\documentclass[mnsc,nonblindrev]{informs3} 

\OneAndAHalfSpacedXI

\usepackage{natbib}
\usepackage{dsfont}
\usepackage{bigints}
\usepackage{tikz}
\usetikzlibrary{intersections,decorations.markings}
\usetikzlibrary{shapes}
\usepackage{cases}
\usepackage{csquotes}
\bibpunct[, ]{(}{)}{,}{a}{}{,}%
\def\bibfont{\small}%
\usepackage{url}
\usepackage[]{algorithm2e}
\usepackage{multirow}
\usepackage{listings}
\usepackage{courier}
\usepackage{upquote}
\usepackage[T1]{fontenc}
\usepackage{textcomp}
\usepackage{listings}
\usepackage{color} 
\definecolor{mygreen}{RGB}{28,100,0} 
\definecolor{mylilas}{RGB}{170,55,241}
\lstdefinestyle{customc}{
  upquote=true,
  language=Matlab,
  showstringspaces=false,
  basicstyle=\small\linespread{0.9}\ttfamily,
  keywordstyle=\color{blue},
  commentstyle=\color{mygreen},
  identifierstyle=\color{black},
  stringstyle=\color{mylilas},
  deletekeywords={get, sin, cos, exp, disp, gamma, diff, fprintf, eps, length, title, plot, subplot, legend, conv, max, abs, sqrt, pi, log, linspace, xlabel, axis}
  }
\TheoremsNumberedThrough 
\ECRepeatTheorems

\EquationsNumberedThrough 

\begin{document}



\RUNTITLE{Tat Lung (Ron) Chan}

\TITLE{Hedging and Pricing European-type, Early-Exercise and Discrete Barrier Options using an Algorithm for the Convolution of Legendre Series}
\ARTICLEAUTHORS{%
\AUTHOR{Tat Lung (Ron) Chan}
\AFF{School of Business, University of East London, Water Lane, Stratford, UK, E15 4LZ, \EMAIL{t.l.chan@uel.ac.uk}} 
\AUTHOR{Nicholas Hale}
\AFF{Department of Mathematical Sciences, Stellenbosch University, Stellenbosch, South Africa, \EMAIL{nickhale@sun.ac.za}}
} 

\ABSTRACT{%
This paper applies an algorithm for the convolution of compactly supported Legendre series (the CONLeg method) \citep[cf.][]{Hal_Tow:2014}, to pricing/hedging European-type, early-exercise and discrete-monitored barrier options under a L\'evy process. The paper employs Chebfun \citep[cf.][]{Tre_Dis:2014} in computational finance and provides a quadrature-free approach by applying the Chebyshev series in financial modelling. A significant advantage of using the CONLeg method is to formulate option pricing and option Greek \textsl{curves} rather than individual prices/values. Moreover, the CONLeg method can yield high accuracy in option pricing and hedging when the risk-free smooth probability density function (PDF) is smooth/non-smooth. Finally, we show that our method can accurately price/hedge options deep in/out of the money and with very long/short maturities. Compared with existing techniques, the CONLeg method performs either favourably or comparably in numerical experiments.

}%


\HISTORY{This paper was first submitted on the 12th November, 2018.}

\maketitle

%

\section{Introduction}\label{sec:intro}

Applying robust numerical techniques in option pricing/hedging and model calibration provides interesting research questions in financial markets. The techniques must be not only highly accurate, but also efficient. 

Suppose we consider the well-known European vanilla option pricing formula driven by a stochastic stock price process $(S_t)_{t\geq 0}:$ 
\begin{align}
\small
V(x,K,t)=e^{-r(T-t)}\mathds{E}(U(S_T,K)\vert S_t=e^x)=e^{-r(T-t)}\int_{-\infty}^{+\infty} U(e^{x+\chi},K) g(\chi)\mathrm{d} \chi,
\end{align}
where $V$ denotes the option value at an initial date of $t$ and with a strike price of $K,$ $U$ stands for a payoff function at the maturity of $T,$ $\mathds{E}$ is the expectation operator under the risk-neutral measure, $x$ and $\chi$ are the log-price and state variable respectively, $S_T$ can be decomposed into $S_t=e^x$ and $e^\chi,$ $g$ is the probability density function of the process, and finally $r$ is a risk-neutral interest rate. 

In the aforementioned formula, $V$ can be seen as a convolution integral, more precisely a cross-correlation integral, and the fast Fourier transform method (FFT method), a numerical integration-based method, \cite[e.g.,][]{Car_Mad:1999, Lew:2001, Lip:2002, Cho:2004, Jac_Jai_Sur:2008, Lor_Fan:2008}, is a popular method for pricing European vanilla options as well as more exotic options, such as the American option, under L\'evy processes. This is because the characteristic function of the underlying dynamics can be easily transformed into a risk-free probability density function (PDF) via the FFT method. The seminal work of these papers leads to the extension of combining the FFT with other transformation methods, e.g., the Hilbert transform or Gaussian transform, in pricing exotic options under the (time-changed) L\'evy process or stochastic volatility models \citep[e.g.,][]{Bro_Yam:2003, Bro_Yama:2005, Fen_Vad:2008, Cai_Kou:2011, Won_Gua:2011, Zen_Kwok:2014}. Within the same numerical integration-based framework, numerical quadrature methods, such as Gaussian quadrature, are proposed mainly by \citet{And_New:2003,Sul:2005,And_New:2007, Che_New:2014} and \citet{Su_New:2017}. The authors of these papers abbreviate their quadrature techniques as QUAD methods. The original QUAD method was introduced in \citet{And_New:2003,And_New:2007} and requires the transition density to be known in closed form, which is the case in, e.g., the Black-Scholes model and Merton's jump-diffusion model. This requirement is relaxed in \citet{Che_New:2014} where the QUAD-FFT or QUAD-CONV method is proposed. The main idea is that the PDF can be recovered by inverting the characteristic function via the FFT method. This helps open up the QUAD method to a much wider range of models. The latest development of the method \citep{Su_New:2017} is to improve the calculation speed, precomputing and caching PDFs and then applying the extrapolation and smoothing techniques to the method. Under the general framework of the QUAD method, only Delta, an option Greek, is formulated via the first-order finite difference method (FD) \citep{And_New:2003}. However, the accuracy of the first-order FD is debatable in the context. Furthermore, other kinds of the Greeks, e.g. Gamma or Theta, are not mentioned or developed yet in this literature. In addition to the QUAD methods, in recent years, \citet{Pac:2018} has introduced the CHEB method, Clenshaw-Curtis quadrature based on an expansion of the integrand in terms of Chebyshev polynomials, for approximate European options with arbitrary payoffs. The method is one of the natural applications of Chebfun \citep{Tre_Dis:2014}, an open-source software system for numerical computing with functions. 

Beyond the FFT method and the QUAD method, Oosterlee and his collaborators have attracted considerable attention \citep{Lee_Oos:2008,Fan_Oos:2009a,Fan_Oos:2009b,Fan_Oos:2011,Zha_Oos:2013,Ruj_Oos:2013}. In their work, they adopt the Fourier cosine series (COS) to price options or derivatives that have different contingency claims and are characterised by path dependence and/or early-exercise features. The implementation of these methods is relatively simple but elegant and is capable of pricing options under different stochastic processes as long as their characteristic function exists. The main achievement of these methods is that they can, in many cases (such as European options), maintain an exponential convergence rate when pricing options. Moreover, these methods are also able to accurately price options under infinite variation processes. The COS method requires an adequate computational domain a priori, and, due to the recursion in time, errors caused by an inadequate domain propagate, resulting in incorrect option prices in the COS method. Moreover, based on the framework of the COS method, Oosterlee and his collaborators further apply the wavelet method called the Shannon wavelet inverse Fourier technique (SWIFT) method to price European, spread, path-dependent and discrete barrier options under exponential L\'evy dynamics \citep[e.g.,][]{Gra_Oos:2013, Gra_Oos:2016,Pap_Oos:2016}. The SWIFT method is used to circumvent the ineffectiveness of using the COS method to price early-exercise options. Nevertheless, comparing with the COS method, a theoretical proof is still lacking to show exponential convergence for the SWIFT method since the wavelet scale $m,$ a parameter to adjust the accuracy of the method, is still chosen heuristically for achieving exponential convergence. Furthermore, the literature on SWIFT methods does not consider computation of Greeks.

In this manuscript we propose the CONLeg method, which uses Chebyshev and Legendre series to approximate a convolution, to improve the ineffectiveness of using the aforementioned methods. First, as mentioned in \citet{Fan_Oos:2009a,Fan_Oos:2009b} and \citet{Chan:2018}, the FFT-driven methods, like the CONV method \citep{Lor_Fan:2008}, are computationally expensive to price option prices or to approximate the PDFs because a relatively large number of Fourier terms is required to obtain sufficient accuracy (see Table \ref{table:VG_ConLeg_individuals} in Section \ref{sec:results}). Second, we question whether any kind of quadrature method is an effective method to approximate option prices/hedging values when option/hedging pricing formulae are treated as convolution integrals. \citet[Section 2 and Section 6,][]{Hal_Tow:2014a} suggest that since a convolution integral appears as a trapezoid (cf. Fig \ref{fig:convLeg}), more quadrature weights and abscissae are required to approximate the areas towards the left and right vertices of the trapezoid. Third, we want to provide a method that can work with any stochastic process with or without a closed-form PDF. For example, the COS and SWIFT methods only work very well when the process has a characteristic function and without applying the FFT, the QUAD method only works well with a closed-form PDF. Fourth, comparing against the CONV, COS, QUAD and SWIFT methods, the CONLeg method provides an option pricing/hedging curve rather than an individual point value. Fifth, unlike the COS method, we can require a proper computational domain a priori in the CONLeg method when we apply it to price early-exercise options. Hence, the CONLeg method will not produce incorrect option prices when we calculate the option prices recursively backwards in time, as we allow a sufficient computational domain for calculation. Sixth, through our numerical experiments, we show that the CONLeg method can achieve high accuracy in option pricing and hedging when the risk-free smooth probability density function (PDF) is non-smooth, and can be an effective method to price/hedge options for long/short maturities. Seventh, to fit in the Chebfun framework, we lay out a closed-form transformation of the complex Fourier expression of a smooth PDF into Chebyshev series when the closed-form PDF is not available (cf. Appendix \ref{sec:CFS_Cheb}). In the similar fashion, we also provide a solution, the Fourier--Pad\'e method (cf. Appendix \ref{sec:sing}), for approximating non-smooth PDFs. Eighth, through this paper, we are keen to promote the convenience of Chebfun \citep{Tre_Dis:2014} in financial modelling. Chebfun is a robust, open source MATLAB pack for computing with functions to 15 digits of accuracy. It contains several state of the art algorithms for Chebyshev and other orthogonal polynomials. Finally, the current CONLeg method is different to the previous literature prompted Chebyshev series and interpolants in option pricing/hedging \citep[cf.][]{Ga:2018,Pac:2018}. As we have mentioned in the previous point, comparing with the CHEB method \citep{Pac:2018}, the CONLeg method is quadrature-free and not limited in pricing/hedging European-type options. Also, unlike \citet{Ga:2018}'s tensorized Chebyshev interpolation to computing Parametric Option Prices (POP), our method does not approximate option pricing curves which must be first precomputed by any numerical method, such as the Monte Carlo and the FFT methods, for some fixed parameter configurations, and then compute other option prices for arbitrary parameter constellations. On the contrary, for a set of fixed parameters, the CONLeg method directly generates an option pricing/hedging curve via approximating a convolution integral without applying other numerical methods for computing the integral first.

The remainder of this paper is structured as follows. Section \ref{sec:ConvoLeg} describes the algorithm for the convolution of Legendre series and how Legendre series can be efficiently computed using the relationship with Chebyshev series and the FFT. Section \ref{sec:Levy_SV} introduces the financial stochastic models we examine in this paper. Section \ref{sec:OptionPriceHedge} describes the formulation of the CONLeg option pricing/Greek formulae for different styles of European options as well as Bermuda, American, and discrete monitored barrier options. Section \ref{sec:trunc} describes the choice of truncated integration intervals. Section \ref{sec:results} discusses, analyses, and compares the numerical results of the CONLeg method with those of other numerical methods discussed above. Finally, we conclude and discuss possible future developments in Section \ref{sec:conclusion}.

\section{Convolution of Legendre Series}\label{sec:ConvoLeg}
Convolution is a fundamental operation that arises in many fields, particularly in financial derivatives research \citep[cf.][]{Car_Mad:1999, Lew:2001, Lip:2002, Jac_Jai_Sur:2008, Lor_Fan:2008}, econometrics \citep{Bon:2003, Liu_Han:2016} and statistics \citep{Hog_Crag:2004}. Given two integrable functions, $f$ and $g$, their convolution is a third function, $h$, defined formally
by the integral
\begin{align}
h(x)=(f\ast g)(x)=\int_{-\infty}^{+\infty} f(y)g(x-y) \mathrm{d}y.
\end{align}
In general, if both $f$ and $g$ are analytic (smooth) and periodic functions, the FFT method, which utilises the convolution theorem and the fast Fourier transform (FFT), is the best choice for approximating $h$ as the FFT approximations of $f$ and $g$ do not suffer the Gibbs phenomenon.  

\begin{figure}[!htb] \label{fig:convLeg}
\centering
\begin{tikzpicture}
\centering
\coordinate (A) at (0, 0);
\coordinate (B) at (3, 3);
\coordinate (C) at (6, 3);
\coordinate (D) at (3, 0);
\coordinate (A1) at (0, -0.3);
\coordinate (B1) at (3, 3.3);
\coordinate (C1) at (6, 3.3);
\coordinate (D1) at (3, -0.3);
\coordinate (E1) at (6, -0.3);
\coordinate (A2) at (-0.5, 0);
\coordinate (B2) at (-0.5, 3);
\coordinate (A3) at (0.5, 2.2);
\coordinate (B3) at (0.5, 2.7);
\coordinate (C3) at (1,2.2);

\draw (A) node[right] {$$} --(B) node[right] {$$};
\draw (B) node[right] {$$} --(C) node[right] {$$};
\draw (C) node[right] {$$} --(D) node[right] {$$};
\draw (D) node[right] {$$} --(A) node[right] {$$};
\draw[densely dotted] (B1) node[right] {$$} -- (D1) node[right] {$c+d$};
\draw (A1) node[right] {$2c$};
\draw (E1) node[right] {$2d$};
\draw (A2) node[right] {$c$};
\draw (B2) node[right] {$d$};
\draw[thick, ->] (A3) node[right] {$$} -- (B3) node[right] {$y$};
\draw[thick, ->] (A3) node[right] {$$} -- (C3) node[right] {$x$};
\end{tikzpicture}
\caption{The convolution domain for two Legendre series on $[ c, d].$}
\end{figure}
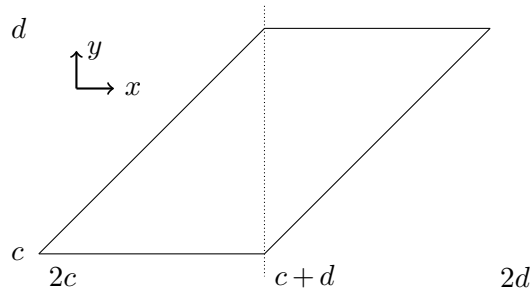

If we now consider $f$ and $g$ $: [c,d]\rightarrow \mathds{R}$ as two compactly supported outside of $[c,d],$  then the convolution aforementioned $h = f\ast g$ via the integral is given by
\begin{align}\label{eqn:conv1}
h(x)=(f\ast g)(x)=\int_{\min(d,x-c)}^{\max(c,x-d)} f(y)g(x-y) \mathrm{d}y,\quad x\in[2c,2d],
\end{align}
and $h(x)=0$ for $x\notin [2c, 2d]$ \footnote{Since convolution is a commutative operation, we consider that only $f$ and $g$ are in the same intervals.}. Without losing any generality, we can visualise each value of $x$ by the diagram in Figure \ref{fig:convLeg}, and we split $h$ into the two pieces suggested by the diagram:
\begin{align}\label{eqn:conv_LR}
h(x)=\begin{cases}h^L(x)=\bigintsss_c^{x-c}f(y)g(y-x) \mathrm{d}y \quad\quad &x\in[2c,c+d],\\ & \\ h^R(x)=\bigintsss_{x-d}^d f(y)g(y-x) \mathrm{d}y \quad\quad &x\in[c+d,2d].\end{cases}
\end{align}
Here, we denote, once and for all, $L$ and $R$ as the \textsl{left} and \textsl{right} hand side of the convolution, respectively.

Based in (\ref{eqn:conv1}), either $f$ or $g$ can be a non-periodic continuous function, the FFT approximations of $f$ and $g$ suffer the Gibbs phenomenon such that there is a permanent oscillatory overshoot in the neighbourhoods of the endpoints $c$ and $d.$ Accordingly, to avoid the Gibbs phenomenon, we adopt the algorithm proposed by \cite{Hal_Tow:2014} to approximate $h.$ The crucial idea of the algorithm is to approximate $f$ and $g$ with finite Legendre series and then convolve the approximations using the convolution theorem for them. The result is a piecewise polynomial representation that can be evaluated at any $x$ in the domain of $h$ to yield an approximation to $h(x)$. If the polynomials used to approximate $f$ and $g$ have degree at most $N$, their algorithm produces an approximation to $h$ in $\mathcal{O} (N^2)$ operations. We summarise the approach below.

To illustrate Hale and Townsends' algorithm of a convolution of two Legendre series, we first define the Legendre series on $[-1,1]$ and then generalise to intervals $[c, d].$  Legendre polynomials, invented by Adrien-Marie Legendre, are the polynomial solutions $P_n(x)$ to Legendre's differential equation
\begin{align}
\frac{\mathrm{d} }{\mathrm{d} x}\left[\left(1-x^2\right) \frac{\mathrm{d} P_n(x)}{\mathrm{d} x}\right]+n(n+1),\quad x\in[-1,1],
\end{align}
with $P_0(x) = 0,$ $P_1(x) = 0$ and integer parameter $n\geq 0.$ $P_n(x)$ forms a polynomial sequence of orthogonal polynomials of degree $n$ and it can be expressed through Rodrigues' formula:
\begin{align}
P_n(x) = \frac{1}{2^n n!} \frac{\mathrm{d}^n}{\mathrm{d}x^n} \left(x^2 -1\right)^n.
\end{align}

Suppose we have $f_M$ and $g_N$ as two finite Legendre series on $[-1,1]$ of degrees $M$ and $N$ with coefficients of $\alpha_0,\ldots,\alpha_M$ and $\beta_0,\ldots\beta_N,$ then we may write $f_M$ and $g_N$ as 
\begin{align}
f_M(x)=\sum\limits_{m=0}^M \alpha_m P_m(x),\quad g_N(x)=\sum\limits_{n=0}^N \beta_n P_N(x).
\end{align} 
With $[c,d]$ equal to $[-1,1],$ the convolution in $(\ref{eqn:conv1})$ becomes
\begin{align}\label{eqn:convo2}
h(x)=(f_M\ast g_N)(x)=\int_{\max(-1,x-1)}^{\min(1,x+1)} f_M(y)g_N(x-y) \mathrm{d}y,\quad x\in[-2,2].
\end{align}
From (\ref{eqn:conv_LR}), $h$, consists of two pieces, $h^L$ on the left with $[-2, 0]$ and $h^R$ on the right with $[0, 2]$, each of degree $N+M+1.$ We construct $h^L$ and $h^R$ by computing their Legendre coefficients. We focus on the computation of $h^L$ since that of $h^R$ is similar. 

Denote by $\{\gamma_k^L\}_k^{M+N+1}$ the vector of the Legendre coefficients of $h^L,$ such that
\begin{align}
h^L(x)=\int_{-1}^{x+1} f_M(y)g_N(x-y)\mathrm{d} y=\sum\limits_{k=0}^{M+N+1} \gamma_k^L P_k(x+1),\quad x\in[-2,0].
\end{align}
By the orthogonality of Legendre polynomials and the orthonormalisation constant $(k + 1/2)^{-1/2}$ for $P_k(x)$ for $k=0,\ldots, M+N+1,$ we have
for $P_k(x)$ for $k = 0, . . . , M + N + 1,$ that
\begin{align}
\gamma_k^L&={2k+1\over 2}\int_{-2}^0 P_k(x+1)\int_{-2}^0 P_k(x+1)\int_{-1}^{x+1} f_M(y)g_N(x-y) \mathrm{d} y \mathrm{d}x\\
&=\sum_{n=0}^{N}\beta_n\underbrace{\left[ {2k+1\over 2} \sum_{m=0}^{M} \alpha_m\int_{-2}^0 P_k(x+1)\int_{-1}^{x+1} P_m(y)P_n(x-y) \mathrm{d} y \mathrm{d}x\right]}_{=B_{k,n}^L}.
\end{align}
\cite{Hal_Tow:2014} prove that the above relation can be expressed in matrix form as $\underline{\gamma}=B^L\underline{\beta}.$ Importantly (\citealt[Theorem 4.1]{Hal_Tow:2014}), there is a three-term recurrence relation of $B_{k,n}^L$ such that 
\begin{align}
B_{k,n}^L&=-{2n+1\over 2k+3}B_{k,n+1}^L+{2n+1\over 2k-1}B_{k,n+1}^L+B_{k,n-1}^L,\quad n,k\geq 1,\\
B_{k,1}^L&=\begin{cases}B_{k-1,0}^L/(2k-1)-B_{k,0}^L-B_{k+1,0}^L/(2k+3),\quad k\neq 0,\\ -B_{1,0}^L/3, \quad k=0\end{cases},\\
B_{k,0}^L&=\begin{cases}{\alpha_{k-1}\over 2k-1}-{\alpha_{k+1}\over 2k+3}, \quad k\neq 0,\\ \alpha_0-\alpha_1/3 \quad k=0 \end{cases}, 
\end{align}
where $0\leq k \leq M + N + 1$ and $0 \leq n \leq N.$ both $B^L_{:,0}$ and $B^L_{:,1}$ can be computed in $\mathcal{O}(M + N)$ operations,
and the whole $(M + N)\times N$ matrix, $B^L$, in $\mathcal{O}((M + N)N)$ operations. The matrix-vector product $B^L\underline{\beta}$ can be computed with the same cost and, accordingly, the
coefficients $\gamma^L$ of $h^L$ in $\mathcal{O}((M + N)N)$ operations. The coefficients $\gamma^R$ of $h^R$ can be computed from $B^R\underline{\beta}$, for which a nearly identical recurrence relation can be derived left since the computation for $h^R$ is similar. 

Now, we direct our attention to the convolution of two finite Legendre series defined on the same interval $[c,d].$ We can define the composition of $P_k\circ\psi_{[c,d]},$ where $\psi_{[c,d]}(x)=(2x-(d+c))/(d-c)$ is the linear mapping from $[c,d]$ to $[-1,1].$ Apart from this, the Legendre series of $f_M$ and $g_N$ on
$[c, d],$ respectively, are formulated by 
\begin{align}\label{eqn:legExt}
f_M(x)=\sum\limits_{m=0}^M \alpha_m P_m\circ\psi_{[c,d]}(x),\quad g_N(x)=\sum\limits_{n=0}^N \alpha_n P_n\circ\psi_{[c,d]}(x)
\end{align}
\citep[Lemma 4.2]{Hal_Tow:2014}. The convolution of $(f\ast g) (x)$ of two continuous functions of $f$ and $g$ defined on $[c, d]$ can be computed as 
\begin{align}
(f\ast g)(x)=\int_{\min(d,x-c)}^{\max(c,x-d)} f(y)g(x-y) \mathrm{d}y={d-c\over 2}\left((f\circ\psi^{-1}_{[c,d]})\ast (g\circ\psi^{-1}_{[c,d]})\right)(y),
\end{align}
where $x\in [2c, 2d]$ and $y=2\psi_{[2c,2d]}(x)\in[-2,2].$

The approach above is not restricted to $f$ and $g$ on the same intervals. In contrast, the algorithm allows $f$ and $g$ which are polynomials of finite degree with zero support outside of $[a, b]$ and $[c, d],$ respectively. In this case (\ref{eqn:conv1}) becomes
\begin{align}\label{eqn:conv2}
h(x)=(f\ast g)(x)=\int_{\min(b,x-c)}^{\max(a,x-d)} f(y)g(x-y) \mathrm{d}y,\quad x\in[a+c,b+d].
\end{align}
Depending on the difference between the length of $[a,b]$ and $[c,d],$ these are three different ideas to compute (\ref{eqn:conv2}) on these subintervals of $[a+c,b+d]$. For the full detail of computing $h$ on these subintervals, we refer interested readers to \citet[Section 5]{Hal_Tow:2014}. The algorithm of convolution of Legendre series supported on same/general intervals is fully implemented as {\tt conv} in Chebfun\footnote{The algorithm in ${\tt conv}$ actually computes the convolution between two Chebyshev series by using fast Chebyshev-Legendre transform \citep{Tow_Web:2018} implemented in {\tt cheb2leg}.}.

\section{L\'evy Processes}\label{sec:Levy_SV}
In this section, we briefly introduce the important properties of one-dimensional L\'evy processes. Standard references for the stochastic processes can be found in \citet{Sch:2003} and \citet{Con_Tan:2004}. Since markets are frictionless and have no arbitrage, we assume that an equivalent martingale measure (EMM) $\mathds{Q}$ is chosen by the market. Moreover, there is a complete filtered probability space $\left(\Omega, \mathcal{F}, \{\mathcal{F}\}_{t\geq 0}, \mathds{Q}\right)$ on which all processes are assumed to live. 

\subsection{L\'evy processes}
With $r\geq 0$ and $q\geq 0$ as the constant risk-free interest rate and the constant dividend yield, respectively, we describe a stock process $(S_t)_{t\geq 0}$ driven by an exponential L\'evy process $(X_t)_{t\geq 0}$ such that
$$S_t=S_0e^{X_t},$$ 
where $X_0=0$ and $X_t$ has infinitely divisible marginal distributions. Given a random variable $X_t,$ we can define it as the corresponding
characteristic function as follows:
\begin{align}\label{eqn:LevyChar}
\varphi(u)&=\mathbb{E}[e^{iuX_t}]=e^{t\phi(u)},\quad u\in \mathbb{R}.
\end{align}
If we define a truncation function $h(\chi) = \chi \mathds{1}_{\vert \chi \vert\leq 1}$ which is a measurable function such that for every $u\in\mathbb{R},$ $\int\vert 1-e^{iu\chi}+iu h(\chi) \vert \nu(\mathrm{d} \chi)<\infty,$ the characteristic function of $X_t$ can be described by the L\'evy--Khinchine representation such that
\begin{align}\label{eqn:LevyKhin}
\phi(u)=iu(r-q+\omega)t-{1\over 2}\sigma^2u^2 + \int_{-\infty}^{+\infty}\left(e^{iu\chi}-1-iu h(\chi) \right)\nu(\mathrm{d} \chi), \quad\chi\in X_{T-t},
\end{align}
Here, $\sigma^2\geq 0$ and $\nu$ are L\'evy measures on $[-\infty, \infty]$ which do not depend on the choice of $h$ (but note that $r-q+\omega$ depends on the choice of it). The condition that $(S_te^{-(r-q)t})_{t\geq 0}$ is a martingale will be guaranteed as long as an appropriate choice of the mean-correcting compensator $\omega$ is calculated as follows:
\begin{align}
\omega&=
{1\over t}\log\varphi(-i)-(r-q).
\end{align}
There are a substantial number of L\'evy process examples in financial modelling. In this paper, we focus on geometric Brownian motion (GBM), variance gamma (VG), normal inverse Gaussian (NIG) and the Carr--Geman--Madan--Yor process (CGMY). Their characteristic functions $\varphi(u)$ can be defined as follows:
\begin{flalign}\label{eqn:charLevyPro}
\small
&\varphi_{\rm GBM}(u):=\exp\left(t\left(iu(r-q+\omega)-\frac{1}{2}\sigma^2u^2\right)\right),\\
&\varphi_{\rm NIG}(u):=\exp\bigg(t\bigg( iu(r-q+\omega) -\frac{1}{2}\sigma^2u^2+\delta\left(\sqrt{\alpha^2-\beta^2}-\sqrt{\alpha^2-(\beta+iu)^2}\right)\bigg)\bigg),\\
&\varphi_{\rm VG}(u):=\exp\bigg(iu(r-q+\omega)t\bigg) \left(\frac{1}{1-i\theta\upsilon u+\frac{\sigma^2\upsilon}{2}u^2}\right)^\frac{t}{\upsilon}, \\
&\varphi_{\rm CGMY}(u):= \exp\Bigg(t\Bigg(iu(r-q+\omega)+C\Gamma(-Y)G^Y\left(\left(1+\frac{iz}{G}\right)^Y-1-\frac{izY}{G}\right)\nonumber\\
&\quad+C\Gamma(-Y)M^Y\left(\left(1-\frac{iz}{M}\right)^Y-1+\frac{izY}{M}\right)\Bigg)\Bigg).
\end{flalign}
\begin{remark}
Most L\'evy processes do not have a closed-form representation of their risk-natural probability density function (PDF) $g$; however, some of the processes, e.g. GBM, NIG and VG, have a closed-formed PDF given by:
\begin{flalign}
g(x)&={1\over \sqrt{2\pi t}\sigma}e^{-{1\over2}{(x-(r-q+\omega_{GBM}))^2\over \sigma^2 t}},\label{eqn:GBM_PDF} \\
g(x)&=\frac{\alpha\delta t K_1 \left(\alpha\sqrt{\delta^2t^2 + (x - (r-q+\omega_{NIG}) t)^2}\right)}{\pi \sqrt{\delta^2t^2 + (x - (r-q+\omega_{NIG}) t)^2}} \; e^{\delta t\sqrt{\alpha^2-\beta^2} + \beta (x - (r-q+\omega_{NIG}) t)},\label{eqn:NIG_PDF} \\
g(x)&={2e^{\theta Z(x) /\sigma^2}\over \nu^{t\over \nu}\sqrt{2\pi}\sigma\Gamma(t/\nu)}\left({Z(x)^2 \over 2\sigma^2/\nu+\theta.^2)}\right)^{{t \over 2\nu}-{1\over 4}}\times K_{{t \over \nu}-{1\over 2}}\left({1\over \sigma^2}\sqrt{Z(x)^2\left({2\sigma^2\over \nu}+\theta^2\right)}\right) \label{eqn:VG_PDF} 
\end{flalign}
respectively. Here, $\omega_{GBM}=-1/2\sigma^2,$ $\omega_{NIG}=\delta\left(\sqrt{\alpha^2-(\beta+1)^2}-\sqrt{\alpha^2-\beta^2}\right),$ $K(\cdot)$ is a modified Bessel function of a second kind, $\Gamma(\cdot)$ is a gamma function, $Z(x)=x-(r-q+\omega_{VG})t$, and $ \omega_{VG}=-1/\nu\log\left(1-\theta\nu-{\sigma^2\nu \over 2}\right).$
\end{remark}

\section{Pricing and Hedging Option via Convolution of Legendre Series}\label{sec:OptionPriceHedge}
In this section, we apply the algorithm for the convolution of Legendre series to formulate option pricing/hedging formulas. 
\subsection{Pricing Formulae for European Type Options}\label{sec:OptionFormu}
Given the current log price $x := \log S,$ the strike price of $K$ and maturity $T \geq t,$ and the probability density function (PDF) $g$ of a stochastic process, we can express the option price $V(x,K,t)$ starting at time $t$ with its contingent claim paying out $U(S_T, K)$ as follows:
\begin{align}\label{eqn:GEquation_1}
V(x,K, t)&=e^{-r (T - t ) } \mathbb{E } (U(S_T, K) \vert S_t = e^x )\nonumber\\
&=e^{-r (T - t ) } \mathbb{E } (U(S_te^{X_T-X_t}, K)\nonumber)\\
&=e^{-r(T-t)}\int_{-\infty}^{+\infty} U(e^{x + \chi-\log K},K) g(\chi) d\chi,\quad \chi \in X_T-X_t=X_{T-t}.
\end{align}
By replacing $x+\chi-\log K$ with $y$, we have 
\begin{align}\label{eqn:GEquation_2}
V(x,K,t)&=e^{-r(T-t)}\int_{-\infty}^{+\infty} U(e^{y},K) g\left(y-x+\log K \right) \mathrm{d}y\nonumber\\
&=e^{-r(T-t)} K\int_{-\infty}^{+\infty} f(y) g^R \left(\tilde{x} - y \right) \mathrm{d}y,
\end{align}
where $\tilde{x}=x-\log K,$ $Kf(y):=U(e^{y},K )$ is the pay-off in log-price coordinates and $g^R (\tilde{x}) := g(-\tilde{x}) $ is the reflected PDF function. 

If $f(y)$ is a piecewise continuous function, as is standard for most options, e.g., vanilla put and call, then applying the FFT method for a convolution function\footnote{The expression of $\int\limits_{-\infty}^{+\infty} U(e^{y},K) g\left(y-x+\log K \right) \mathrm{d}y\nonumber$ in (\ref{eqn:GEquation_1}) is indeed a cross-correlation integral; however, since we introduce the idea of the reflected function $g^R (\tilde{x}) := g(-\tilde{x})$, we can turn (\ref{eqn:GEquation_1}) into a convolution integral instead.}, $(f*g^R) (\tilde{x}) := \int_{-\infty}^{+\infty} f(y ) g^R(\tilde{x} - y) \mathrm{d}y,$ in $[-\infty, \infty]$ will cause Gibbs phenomenon and as a result, will affect the accuracy of approximating $V$. To avoid Gibbs phenomenon and allow a good approximation of $V,$ we replace $[-\infty, \infty]$ with an interval $[c,d]$ and employ the CONLeg method. The choice of $[c,d]$ satisfies the condition of 
\begin{align}\label{eqn:charAppx}
\int_{c}^{d} g(\chi) e^{iu\chi} \rm{d}\chi\approx\int_{-\infty}^{+\infty} g(\chi) e^{iu\chi} \rm{d}\chi=\mathds{E}[e^{iu(X_T-X_t)}]:=\varphi(u), 
\end{align}
where $\varphi(u)$ is a characteristic function of $X_T-X_t.$ Then we can approximate the pricing formula $V$ on $[c,d],$ i.e.,
\begin{align}\label{eqn:GEquation_3}
V(x,K, t)&\approx e^{-r(T-t)}K\int_{c}^{d} f(y)g^R(\tilde{x}-y) \mathrm{d}y.\nonumber\\
&=e^{-r(T-t)}K\int_{\min(d,\tilde{x}-c)}^{\max(c,\tilde{x}-d)} f(y)g^R(\tilde{x}-y) \mathrm{d}y,\quad \tilde{x}\in[2c,2d]\nonumber\\
&=e^{-r(T-t)} Kh(\tilde{x}).
\end{align}
where, $h(\tilde{x})$ has compact support outside of $[2c, 2d].$ The final form of (\ref{eqn:GEquation_3}) is ready for approximating via the CONLeg method.

Most of the closed-form expressions of $g$ do not exist in the stochastic processes. If they do not, we adopt the ideas proposed in \citet{Chan:2016, Chan:2018} to express $g$ in a complex Fourier series (CFS) representation such that $g$ is approximated by:
\begin{align}\label{eqn:CFS_PDF}
g_N(y):=\mathfrak{Re}\left[2\sum_{k=1}^{N} b_k e^{i\frac{2\pi}{d-c}ky}+b_0\right],
\end{align}
where $i$ is a complex number and $\mathfrak{Re}$ is the real part of a complex number, and given the condition of (\ref{eqn:charAppx}), 
\begin{align}\label{eqn:B_k}
b_k=\int_{c}^{d} g(y)e^{-i\frac{ 2\pi }{d-c}ky} \mathrm{d}y\approx\varphi\left(-\frac{2\pi}{d-c}k\right)\quad\hbox{and}\quad b_0=\int_{c}^{d} g(y)\mathrm{d}y\approx\varphi(0)=1. 
\end{align}
For the expression of $g^R,$ we simply put a negative sign in the basis function $e^{i {2\pi \over d-c} ky },$ i.e., 
\begin{align}\label{eqn:CFS_PDF}
g^R(y)\approx g_N^R(y):=\mathfrak{Re}\left[2\sum_{k=1}^{N} b_k e^{-i\frac{2\pi}{d-c}ky}+b_0\right].
\end{align}
If $g^R$ is smooth throughout on $[c,d],$ we can either directly approximate the CFS representation with a Chebyshev series using Chebfun \citep[cf.][]{Tre_Dis:2014} or transform it into a Chebyshev series using the techniques shown in Appendix \ref{sec:CFS_Cheb}. If $g^R$ is a piecewise continuous function containing a singularity\footnote{We refer a singularity as a point at which is not defined, or a point which fails to be well-behaved after differentiability}, then we use the Fourier--Pad\'e ideas to locate the singularity in $g^R$ and form accurate approximation. The details can be found in the Appendix \ref{sec:sing}.

Knowing singularities $\tilde{x}_1\ldots \tilde{x}_{\mathcal{K}+1}$ in $[c,d],$ we divide $f$ and $g^R$ into a set of piecewise continuous functions and then approximate them with Chebyshev series using {\tt chebfun} \citep[cf.][Chapter 1.4]{Tre_Dis:2014}. Accordingly, we have a set of $\mathcal{K}$ polynomials 
$f_M$ and $g_N,$ each of degree at most $M$ and $N$ on the subintervals $[\tilde{x}_k, \tilde{x}_{k+1}],$ i.e., 
\begin{align}
f_M=\sum\limits_{k=1}^\mathcal{K} f_{k, M}\mathds{1}_{[\tilde{x}_k, \tilde{x}_{k+1}]},\quad g^R_N=\sum\limits_{k=1}^\mathcal{K} g^R_{k,N}\mathds{1}_{[\tilde{x}_k, \tilde{x}_{k+1}]}.
\end{align}
Here, $\mathds{1}_{[\tilde{x}_k, \tilde{x}_{k+1}]}$ as the indicator function in the interval $[\tilde{x}_k, \tilde{x}_{k+1}]$ and
\begin{align}
f_{k,M}(\tilde{x})=\sum\limits_{m=0}^M\alpha_{k,m}^{cheb} T_m\circ\psi_{[\tilde{x}_k, \tilde{x}_{k+1}]}(\tilde{x}),\quad g^R_{k,N}(x)=\sum\limits_{n=0}^N\beta_{k,n}^{cheb} T_n\circ\psi_{[\tilde{x}_k, \tilde{x}_{k+1}]}(\tilde{x}). 
\end{align}
Then, using the techniques implemented in {\tt cheb2leg} \citep[cf.][]{Tow_Web:2018}, we transform both $f_N$ and $g^R_M$ into Legendre series defined by 
\begin{align}
f_{k,M}(\tilde{x})=\sum\limits_{m=0}^M\alpha_{k,m}^{leg} P_m\circ\psi_{[\tilde{x}_k, \tilde{x}_{k+1}]}(\tilde{x}),\quad g^R_{k,N}(\tilde{x})=\sum\limits_{n=0}^N\beta_{k,n}^{leg} P_{n}\circ\psi_{[\tilde{x}_k, \tilde{x}_{k+1}]}.
\end{align}
We use the algorithm for the convolution of Legendre series described in Section \ref{sec:ConvoLeg} to approximate their convolution $h(\tilde{x})=(f_M*g_N)(\tilde{x})$ on the subintervals $[\tilde{x}_k, \tilde{x}_{k+1}].$ Finally, transforming $h(\tilde{x})$ back into Chebyshev series using {\tt leg2cheb}, $V$ can be approximated by
\begin{align}
e^{-r(T-t)}Kh(\tilde{x})&=e^{-r(T-t)}K\left(f_M*g_N\right)(\tilde{x})\label{eqn:final_LegCheForm}\\
&=e^{-r(T-t)}K\sum\limits_{k=1}^{M_k} V_{N_k}\mathds{1}_{[\tilde{x}_k, \tilde{x}_{k+1}]}
\intertext{where,}
V_{N_k}&=\sum\limits_{k=1}^{N_k}\gamma_k T_{k}\circ\psi_{[\tilde{x}_k, \tilde{x}_{k+1}]}(\tilde{x})
\end{align}
Using (\ref{eqn:final_LegCheForm}), we can generate a set of option prices with a value of $K$ and a range of $S_t.$ However, in the financial markets, option price quotes always appear with a value of $S_t$ and a range of $K.$ To fit in this financial phenomenon, we modify (\ref{eqn:final_LegCheForm}) using the fact of $K=S e^{-\tilde{x}}=e^{x-\tilde{x}}$ so that we obtain the new pricing formula of
\begin{align}\label{eqn:final_LegCheForm1}
V(x,K, t)=e^{-r(T-t)+x-\tilde{x}}h(\tilde{x}).
\end{align}
\begin{remark}
In Chebfun, there is a built-in algorithm to detect singularities automatically in a piecewise continuous function \citep[cf.][]{Pac_Tre:2010}. However, since PDFs can be easily expressed in the complex Fourier series and then extend to the Fourier--Pad\'e series \citep[cf.][]{Chan:2016, Chan:2018}, we instead use Fourier--Pad\'e ideas to locate singularities or approximate PDFs. 
\end{remark}

\subsubsection{European Vanilla Call Options as illustration} 
We now consider pricing a European vanilla call, which can be exercised only at its maturity, defined in (\ref{eqn:GEquation_1}) with a payoff function of 
\begin{align}
U(S_T,K)=\max\left(S_T-K\right).
\end{align}
We first transform the payoff into
\begin{align}
\max\left(e^{x+\chi}-K,0\right)=K\max\left(e^{x+\chi-\log K}-1,0\right). 
\end{align}
By replacing $x+\chi-\log K$ with $y$, we have a new form of $V(x,K,t)$ denoted as
\begin{align}
V(x,K,t)&=e^{-r(T-t)}K\int_{-\infty}^{\infty} \max(e^{y}-1,0) g\left(y-x+\log K\right) \mathrm{d}y \nonumber\\
&=e^{-r(T-t)}K\int_{-\infty}^{\infty} \max(e^{y}-1,0) g^R\left(\tilde{x}-y\right) \mathrm{d}y,
\end{align}
where $\tilde{x}=x-\log K$ and $g^R(\tilde{x}):=g(-\tilde{x})$ is a reflecting function.
To make the CONLeg more efficient, we define a truncated computational interval $[c,d]$ (cf. Section \ref{sec:trunc}), which satisfies condition $(\ref{eqn:charAppx}),$ to replace $[-\infty,\infty].$ Then, $V(x,K,t)$ is reformulated as
\begin{align}\label{eqn:EuroV_1}
V(x,K,t)&\approx e^{-r(T-t)}K\int_{c}^{d}\max(e^{y}-1,0) g^R\left(\tilde{x}-y\right) \mathrm{d}y\nonumber\\
&=e^{-r(T-t)}K\int_{\min(d,x-c)}^{\max(c,x-d)} f(y)g^R(\tilde{x}-y) \mathrm{d}y,\quad \tilde{x}\in[2c,2d]\nonumber\\
&=e^{-r(T-t)}Kh(\tilde{x}),
\end{align}
where $f(y):=\max(e^{y}-1,0).$ To easily digest how the CONLeg method approximates $h(\tilde{x})$, we assume that $g^R$ is a piecewise smooth function containing only one jump $y=0$ appearing in $f,$ the payoff function, on $[c,d]$. We can use {\tt chebfun} to approximate $f$ and $g^R$ on $[c,0,d]$\footnote{One should note that when $y\leq 0,$ $\max(e^{y}-1,0)=0$.} and $[c,d]$ respectively. Using the techniques described in Section \ref{sec:OptionFormu}, the European call option pricing formula is given by:
\begin{align}
V(x,K, t)&\approx e^{-r(T-t)}K\left(f_M*g_N\right)(\tilde{x})\nonumber\\
&=e^{-r(T-t)}K\sum\limits_{k=1}^4V_{N_k}\mathds{1}_{[\tilde{x}_k, \tilde{x}_{k+1}]}\quad\tilde{x}\in [2c, 2d],
\end{align}
where, $V_{N_k}=\sum\limits_{k=1}^{N_k}\gamma_k T_k\circ\psi_{[\tilde{x}_k, \tilde{x}_{k+1}]}(\tilde{x}),$ and $\tilde{x}_1=2c, \tilde{x}_2=c, \tilde{x}_3=0, \tilde{x}_4=d, \tilde{x}_5=2d.$ 

Moreover, if we only focus on $[c,d],$  we have
\begin{align}
V(x,K, t)=e^{-r(T-t)}K\left(V_{N_1}\mathds{1}_{[c,0]}+V_{N_2}\mathds{1}_{[0,d]}\right)\quad\tilde{x}\in [c, d],
\end{align}
where $\tilde{x}_1=c, \tilde{x}_2=0, \tilde{x}_3=d.$ See Section \ref{sec:results_Euro} for an example.

The CONLeg method is not limited from pricing European vanilla call option (\ref{eqn:EuroV_1}); it can be readily extended into a put option or other options with different pay-off structures, e.g., Cash-or-Nothing options. In Table \ref{table:CFSPayoff1}, we list all financial contingency claims we consider in this paper and both their payoff functions and transformed payoff functions. 
\begin{table}\label{table:CFSPayoff1}
\caption{Payoff functions and their transforms for a variety of financial contingency claims. $\mathds{1}$ represents an indicator function and $n<\infty$ is any positive integer. The singularity always exits at $y=0$ in the transformed payoff function when $y=x+\chi-\log K.$} 
\centering 
\small\addtolength{\tabcolsep}{-1pt}
\begin{tabular}{|ccc|} 
\hline
\rule{0pt}{3ex}
Financial Contingency Claim&Payoff Function&Transformed Payoff Function\\[10pt]
&$U(S_T,K)$&$U(e^{x+\chi-\log K},K)$\\[10pt]
\hline
\rule{0pt}{4ex} 
Call&$\max(S_T-K,0)$&$K\max(e^{x+\chi-\log K}-1,0)$\\[10pt]
Put&$\max(K-S_T,0)$&$K\max(1-e^{x+\chi-\log K},0)$\\[10pt]
Covered Call&$\min(S_T,K)$&$K\min(e^{x+\chi-\log K}-1,0)+K$\\[10pt]
Cash-or-Nothing Call&$\mathds{1}_{S_T\geq K}$&$\mathds{1}_{e^{x+\chi-\log K}\geq 1}$\\[10pt]
Cash-or-Nothing Put&$\mathds{1}_{S_T\leq K}$&$\mathds{1}_{e^{x+\chi-\log K}\leq 1}$\\[10pt]
Asset-or-Nothing Call&$S_T\mathds{1}_{S_T\geq K}$&$e^{x+\chi}\mathds{1}_{e^{x+\chi-\log K}\geq 1}$\\[10pt]
Asset-or-Nothing Put&$S_T\mathds{1}_{S_T\leq K}$&$e^{x+\chi}\mathds{1}_{e^{x+\chi-\log K}\leq 1}$\\[10pt]
Asymmetric Call&$(S_T^n-K^n)\mathds{1}_{S_T\geq K}$&$K^n(e^{n(x+\chi-\log K)}-1)\mathds{1}_{e^{x+\chi-\log K}\geq 1}$\\[10pt]
Asymmetric Put&$(K^n-S_T^n)\mathds{1}_{S_T\leq K}$&$K^n(1-e^{n(x+\chi-\log K)})\mathds{1}_{e^{x+\chi-\log K}\leq 1}$\\[10pt]
\hline
\end{tabular}
\end{table}
\subsection{Pricing Formulae for Bermuda Options}\label{sec:Bermuda}
Consider now $\log S_t:=x_t$ driven by a L\'evy process and a Bermudan option with strike $K$ and maturity $T$ that can be exercised only on a given number of exercise dates $t=t_0 <t_1\leq t_2\leq\ldots t_l\leq t_{l+1}\leq\ldots\leq t_L = T.$ We can write the Bermudan pricing formula for such an option as
\begin{align}\label{eqn:Bermuda_formulae1}
V(x_{t_l},K,t_l)&=\begin{cases} U(e^{x_{t_l}},K,t_l) & l=L,\, t_L=T\\
\max\left(C(x_{t_l},K,t_l), U(e^{x_{t_l}},K,t_l) \right) & l=1,2,3,\dots,L-1\\
C(x_{t_l},K,t_l)&l=0
\end{cases},
\end{align}
where, $U(e^{x_{t_l}},K,t_l)$ is the payoff function at $t_l.$ That is, if the payoff function is a call, then $U(e^{x_{t_l}},K,t_l)$ is transformed into $\max\left(e^{x_{t_l}}-K,0\right).$ In (\ref{eqn:Bermuda_formulae1}), $C(x_{t_l},K,t_l)$ at each $t_j$ can be defined as
\begin{align}
C(x_{t_j},K,t_j)&=e^{-r(t_{j+1}-t_j)}\mathbb{E}\left(V(x_{t_{j+1}},K,t_{j+1}) \vert x_{t_j} \right).
\end{align}
To apply the CONLeg method to approximate $C(x_{t_l},K, t_l),$ we first understand that 
\begin{align}
S_{t_l}=e^{-r(t_{l+1}-t_l)}\mathbb{E}\left(S_{t_{l+1}}\vert S_{t_l}=e^{x_{t_l}}\right)=e^{-r(t_{l+1}-t_l)}\mathbb{E}\left(e^{x_{t_l}+X_{t_{l+1}}-X_{t_l}}\right)=e^{x_{t_l}}
\end{align}
is a martingale process. We also denote $\tilde{x}_{t_l}$ as $x_{t_l}-\log K$ and follow Section \ref{sec:OptionFormu} to approximate $C(x_{t_l},K,t_l)$ as European option prices at $t_l$. Then we can transform $C(x_{t_l},K,t_l)$ into
\begin{align}\label{eqn:C_tlEuro}
&=e^{-r(t_{l+1}-t_l)}\mathbb{E}\left(V(x_{t_{l+1}},K,t_{l+1}) \vert x_{t_l} \right)\nonumber\\
&=e^{-r(t_{l+1}-t_l)}\int^{+\infty}_{-\infty} V(x_{t_l}+\chi-\log K,t_{l+1})g(\chi)\mathrm{d} \chi,\quad \chi\in X_{t_{l+1}}-X_{t_l}\nonumber\\
&=e^{-r(t_{l+1}-t_l)}K\int_{\min(d,x-c)}^{\max(c,x-d)} f(y)g^R(\tilde{x}_{t_l}-y) \mathrm{d}y\nonumber\\
&=e^{-r(t_{l+1}-t_l)}Kh(\tilde{x}_{t_l}).
\end{align}
Since we approximate $h(\tilde{x}_{t_l})$ with the CONLeg method and use the Chebyshev series to present the Bermuda option prices $C(x_{t_l},K,t_l),$ accordingly, we can further modify (\ref{eqn:Bermuda_formulae1}) with a new form of 
\begin{align}\label{eqn:Bermuda_formulae2}
V(x_{t_l},K,t_l)&=\begin{cases} K\tilde{f}(\tilde{x}_{t_l}) & l=L,\, t_L=T\\
K\max\left(e^{-r(t_{l+1}-t_l)}h(\tilde{x}_{t_l}), \tilde{f}(\tilde{x}_{t_l})\right)& l=1,2,3,\dots,L-1\\
Ke^{-r(t_{l+1}-t_l)}h(\tilde{x}_{t_l})&l=0
\end{cases},
\end{align}
where $K\tilde{f}(\tilde{x}_{t_l}):=U(e^{x_{t_l}},K,t_l).$ Since the no-arbitrage assumption leads to the requirement that $\partial V/\partial x$ is continuous and $V(x_{t_l},K,t_l)=U(e^{x_{t_l}},K,t_l)$ at the early exercise curve, we must determine exercise point $x^*_{t_l}$ appearing in $V(x_{t_l},K,t_l)=U(e^{x_{t_l}},K,t_l).$ One way to do this is to use the Newton method proposed in \citet{Fan_Oos:2009b} to find $x_{t_l}.$ However, since $V$ and $U$ are represented by piecewise smooth polynomials ({\tt chebfun}), we can apply a built-in function {\tt roots} in Chebfun to efficiently find these zeros. To do so, we first approximate $\tilde{f}(\tilde{x}_{t_l})$ as a Chebyshev series, then apply the roots function to find $\tilde{x}^*_{t_l}$ in the following equality:
$$e^{-r(t_{l+1}-t_l)}h(\tilde{x}_{t_l})-\tilde{f}(\tilde{x}_{t_l})=0.$$
Once we have $\tilde{x}^*_{t_l}$ and use it as a break point, we approximate 
\begin{align}
\max\left(e^{-r(t_{l+1}-t_l)}h(\tilde{x}_{t_l}), f(\tilde{x}_{t_l})\right)
\end{align}
with two different Chebyshev series. Considering $\tilde{x}^*_{t_l}$ as a singularity and combining other singularities, $\tilde{x}_{t_l,1}\ldots \tilde{x}_{t_l, \mathcal{K}+1}\in\tilde{x}_{t_l},$ e.g., singularities in a non-smooth PDF and/or in a payoff function, in $V(x_{t_l},K,t_l),$ 
we approximate $V(x_{t_l},K,t_l),$ with a set of Chebyshev series given by
\begin{align}\label{eqn:Ber_Cheb}
K\sum\limits_{k=0}^{M_k} V_{N_k}\mathds{1}_{[\tilde{x}_{t_l,k}, \tilde{x}_{t_l,k+1}]} \hbox{ and } V_{N_k}=\sum\limits_{k=1}^{N_k}\gamma_k T_k\circ\psi_{[\tilde{x}_k, \tilde{x}_{k+1}]}(\tilde{x}).
\end{align}
Finally, summarising the methods above, we present the pseudo-code of our algorithm computing Bermudan option prices in Algorithm \ref{algo:Bermuda}. A numerical example is also presented in Section \ref{sec:results_Exotic}.
\begin{algorithm}[h]
\caption{Algorithm for computing Bermudan option price $V(x_t, K, t)$ at time $t$ based on (\ref{eqn:Bermuda_formulae1}).}\label{algo:Bermuda}
\KwResult{Bermuda option price $V(x_{t},K,t)$ at time t}
initialisation\;
discretise $[t,T]$ into timesteps $t=t_0, t_1,\ldots, t_l,\ldots, t_L=T$\;
$ t_l=t_{L-1}$\;
\While{$t_l\neq t$}{
compute $C(x_{t_l},K, t_l)$ using the CONLeg method\;
$C(x_{t_l},K, t_l)=e^{-r(t_{l+1}-t_l)}h(\tilde{x}_{t_l})$ in (\ref{eqn:Bermuda_formulae2})\;
find $\tilde{x}^*_{t_l}$ in $e^{-r(t_{l+1}-t_l)}h(\tilde{x}_{t_l})-\tilde{f}(\tilde{x}_{t_l})=0$\;
compute $\max\left(e^{-r(t_{l+1}-t_l)}h(\tilde{x}_{t_l}), \tilde{f}(\tilde{x}_{t_l})\right)$ with two Chebyshev series (\ref{eqn:Ber_Cheb})\;
$V(x_{t_l},K, t_l)=K\sum\limits_{k=0}^{M_k} V_{N_k}\mathds{1}_{[\tilde{x}_{t_l,k}, \tilde{x}_{t_l,k+1}]}$\;
next $t_l$\;
}
return $V(x_t,K, t)$ equal to $e^{-r(t_1-t)}Kh(\tilde{x}_t),$ where $t_0=t$\;
\end{algorithm}

\subsection{Pricing Formulae for American Options}
There are two basic approaches to evaluating American options based on our method for Bermudan options. As suggested in \citet{Fan_Oos:2009b}, one simple approach is to approximate an American option by a Bermudan option with many exercise opportunities. In other words, increase the number of exercise opportunities $L$ to a very large value. An alternative approach is to use Richardson extrapolation on a series of Bermudan options with an increasing number of $L$ \citep[cf.][]{Ges_Joh:1984, Cha_Sta:2007}. We adapt the latter approach (which is also implemented in \cite{Fan_Oos:2009b}) to price the American option here. Therefore, implementing the 4-point Richardson extrapolation scheme \citep[cf.][]{Fan_Oos:2009b}, we have the American option price given by 
\begin{align}\label{eqn:amer_extrapolation}
V_{Amer}(L)={1\over 21}\left(64V(2^{L+3})-56V(2^{L+2})+14V(2^{L+1})-V(2^{L})\right),
\end{align}
where $V_{Amer}(L)$ denotes the approximated value of the American
option and $V(\cdot)$ is the pricing formulae for Bermudan options in (\ref{eqn:Bermuda_formulae2}).

\subsection{Pricing Formulae for Discretely Monitored Barrier Options}\label{sec:barrier}
A barrier option is an early-exercise option whose payoff depends on the stock price crossing a pre-set barrier level during the option's lifetime. We call the option an up-and-out, knock-out, or down-and-out option when the option's existence fades out after crossing the barrier level. Like European vanilla options, these options can all be written as either put or call contracts that have a pre-determined strike price on an expiration date. In this paper, we only investigate two basic types of barrier options.
\begin{enumerate}
\item \textsl{Down-and-out barrier (DO) option}: A down-and-out barrier option is an option that can be exercised at a pre-set strike price on an expiration date as long as the stock price that drives the option does not go below a pre-set barrier level during the option's lifetime. As an illustration, if the stock price falls below the barrier, the option is ``knocked-out'' and immediately carries no value. 
\item \textsl{Up-and-out barrier (UO) option}: Similar to a down-and-out barrier option, an up-and-out barrier option will be knocked out when the stock price rises above the barrier level during the option's lifetime. Once it is knocked out, the option cannot be exercised at a predetermined strike price on an expiration date.
\end{enumerate}

The structure of discretely monitored barrier options is the same as that of Bermudan options. Instead of having a pre-set exercise date and an early-exercise point like Bermudan options, barrier options have a pre-set monitored date and a barrier level. In the case of Bermudan options, when the stock price goes across the early exercise point, a payoff occurs, and the option expires immediately. In the same manner, a barrier option knocks out immediately when the barrier level is crossed. The barrier level acts exactly the same as the exercise point in Bermudan options. However, in the case of a barrier option without a rebate, no payoff occurs when the barrier level is reached; otherwise, a rebate occurs when a barrier option is knocked out. 

We use a rebate DO option to illustrate the CONLeg method to approximate discretely monitored barrier option prices. Suppose that we have a rebate DO option driven by $S_t$ with a barrier $B$, a rebate $R_b,$ a strike $K$ and a series of monitoring dates $L$: $t=t_0<\ldots<t_l<\ldots < t_L=T;$ the option formulae can be described as
%
\begin{align}\label{eqn:Barrier_formulae1}
V(x_{t_l},K,t_l)&=\begin{cases} U(e^{x_{t_l}},K,t_l)\mathds{1}_{\log B> x_{t_l}} + R_b\mathds{1}_{\log B \leq x_{t_l}}& l=L,\, t_L=T\\
C(x_{t_l},K,t_l)\mathds{1}_{\log B> x_{t_l}}+e^{-r(T-t_l)}R_b\mathds{1}_{\log B \leq x_{t_l}} & l=1,\dots,L-1\\
C(x_{t_l},K,t_l)&l=0
\end{cases},
\end{align}
where $\mathds{1}$ is an indicator function and $ U(e^{x_{t_l}},K,t_l)$ is again either a call or put payoff. We follow the ideas of (\ref{eqn:C_tlEuro}) and (\ref{eqn:Bermuda_formulae2}) in Section \ref{sec:Bermuda} to approximate $C(x_{t_l},K,t_l)$ such that 
\begin{align}
e^{-r(t_{l+1}-t_l)}\mathbb{E}\left(V(x_{t_{l+1}},K,t_{l+1}) \vert x_{t_l} \right)=e^{-r(t_{l+1}-t_l)}Kh(\tilde{x}_{t_l}).
\end{align}
After we apply the CONLeg method, (\ref{eqn:Barrier_formulae1}) can be transformed into
\begin{align}\label{eqn:Barrier_formulae2}
V(x_{t_l},K,t_l)=\begin{cases} K\left(\tilde{f}(\tilde{x}_{t_l})\mathds{1}_{\log (B/K)> \tilde{x}_{t_l}} + {R_b\over K}\mathds{1}_{\log(B/K) \leq x_{t_l}}\right)& l=L,\, t_L=T\\
K\left(e^{-r(t_{l+1}-t_l)}h(\tilde{x}_{t_l})\mathds{1}_{\log ({B\over K})> \tilde{x}_{t_l}} +e^{-r(T-t_l)}{R_b\over K}\mathds{1}_{\log({B \over K}) \leq x_{t_l}}\right)& l=1,\dots,L-1\\
Ke^{-r(t_{l+1}-t_l)}h(\tilde{x}_{t_l})&l=0
\end{cases}.
\end{align}
In (\ref{eqn:Barrier_formulae2}), since there is a jump at $\log(B/K),$ the barrier, at $t_l$, we use $\log(B/K)$ as a break point and approximate
\begin{align}
e^{-r(t_{l+1}-t_l)}h(\tilde{x}_{t_l})\mathds{1}_{\log ({B\over K})> \tilde{x}_{t_l}} +e^{-r(T-t_l)}{R_b\over K}\mathds{1}_{\log({B \over K}) \leq x_{t_l}}
\end{align}
with two Chebyshev series. Moreover, combining other singularities $\tilde{x}_{t_l,1}\ldots \tilde{x}_{t_l, \mathcal{K}+1}\in\tilde{x}_{t_l}$ in $V(x_{t_l},K,t_l)$, we can formulate $V(x_{t_l},K,t_l)$ with a set of Chebyshev series given in (\ref{eqn:Ber_Cheb}). Finally, the pseudo-code of our algorithm calculating discretely monitored DO barrier option prices can be found in Algorithm \ref{algo:Barrier}.
\begin{algorithm}[h]
\caption{Algorithm for computing discretely monitored barrier option price $V(x_t, K, t)$ at time $t$ based on (\ref{eqn:Barrier_formulae1}).}\label{algo:Barrier}
\KwResult{discretely monitored barrier option price $V(x_{t},K,t)$ at time $t$}
initialisation\;
discretise $[t,T]$ into timesteps $t=t_0, t_1,\ldots, t_l,\ldots, t_L=T$\;
$ t_l=t_{L-1}$\;
\While{$t_l\neq t$}{
compute $C(x_{t_l},K, t_l)$ using the CONLeg method\;
$C(x_{t_l},K, t_l)=e^{-r(t_{l+1}-t_l)}h(\tilde{x}_{t_l})$ in (\ref{eqn:Barrier_formulae2})\;
compute $\begin{cases}e^{-r(t_{l+1}-t_l)}h(\tilde{x}_{t_l})& \mbox{if } \log (B/K)> \tilde{x}_{t_l}\\
e^{-r(T-t_l)}R_b/K &\mbox{if } \log(B/K) \leq \tilde{x}_{t_l} \end{cases}$ with two Chebyshev series (\ref{eqn:Ber_Cheb})\;
$V(x_{t_l},K, t_l)=K\sum\limits_{k=0}^{M_k} V_{N_k}\mathds{1}_{[\tilde{x}_{t_l,k}, \tilde{x}_{t_l,k+1}]}$\;
next $t_l$\;
}
return $V(x_t,K, t)$ equal to $e^{-r(t_1-t)}Kh(\tilde{x}_t),$ where $t_0=t$\;
\end{algorithm}

For the UO barrier options, we can use (\ref{eqn:Barrier_formulae2}) and Algorithm \ref{algo:Barrier} to compute their prices, but we consider the condition of the option knocked out when the stock price rises above $B,$ i.e.,
\begin{align}
V(x_{t_l},K, t_l)=\begin{cases} U(e^{x_{t_l}},K,t_l)\mathds{1}_{\log B< x_{t_l}} + R_b\mathds{1}_{\log B\geq \tilde{x}_{t_l}} & l=L,\, t_L=T\\
C(x_{t_l},K,t_l)\mathds{1}_{\log B< x_{t_l}}+e^{-r(T-t_l)}R_b\mathds{1}_{\log B\geq \tilde{x}_{t_l}}& l=1,\dots,L-1\\
C(x_{t_l},K,t_l)&l=0
\end{cases}.
\end{align}

\subsection{Hedging Formulae and Choice of Truncated Intervals}\label{sec:OptionHedge}
We now turn our attention to deriving the option Greek values. In particular, we focus on deriving three option Greek values---Delta ($\Delta$), Gamma ($\Gamma$), and Vega. Delta is defined as the rate of change in the option value with respect to changes in the underlying asset price; Gamma is the rate of change of Delta with respect to changes in the underlying price; and finally, Vega is the measurement of an option's sensitivity to changes in the volatility of the underlying asset price. In general, volatility measures the amount and speed at which the price moves up and down and is often based on changes in the recent, historical prices of a trading instrument. Other Greek values, such as Theta, can be derived in a similar fashion; however, depending on the characteristic function, the derivation expressions might be rather lengthy. We omit them here, as many terms are repeated. 

As mentioned above, Delta is the first derivative of the value $V$ of the option with respect to the underlying instrument price $S$. Hence, differentiating the convolution form of $V$ in European options (\ref{eqn:final_LegCheForm}), Bermuda options (\ref{eqn:Ber_Cheb}), American options (\ref{eqn:amer_extrapolation}) and barrier options (\ref{eqn:Ber_Cheb}) with respect to $S,$ we have 
\begin{align}\label{eqn:LegCon_delta}
\Delta_t&={\partial V(x,K,t) \over \partial S}=e^{-r(T-t)}K{\partial h(\tilde{x}) \over \partial \tilde{x}}{\partial \tilde{x}\over \partial x}{\partial x\over \partial S},\quad \tilde{x}=x-\log K.
\end{align} 
Since $\partial x / \partial \tilde{x}=1$ and $\partial x / \partial S =\exp(-x),$ $\Delta_t$ simply becomes
\begin{align}
\quad e^{-r(T-t)-x}K{\partial h(\tilde{x}) \over \partial \tilde{x}}&=e^{-r(T-t)-x}K\sum\limits_{k=1}^{M_k} {\partial V_{N_k} \over \partial \tilde{x}} \mathds{1}_{[\tilde{x}_k, \tilde{x}_{k+1}]},
\intertext{with}
 {\partial V_{N_k} \over \partial \tilde{x}}&=\sum\limits_{k=1}^{N_k} \gamma_k {\partial T_k\circ\psi_{[\tilde{x}_k, \tilde{x}_{k+1}]}(\tilde{x}) \over \partial \tilde{x}}.\label{eqn:Cheb1st}
\end{align}
To express the first derivative of the Chebyshev series in (\ref{eqn:Cheb1st}), we adopt the fact of 
$${d \over d x}T_n(x)={n\over 2}{T_{n-1}(x)-T_{n+1}(x)\over 1-x^2}$$\citep[cf.][(2.4.5)]{Mas_Han:2002} and $\mathrm{d} \psi_{[c,d]}(\tilde{x}) =(2/(d-c))\mathrm{d} \tilde{x},$ 
such that we have
\begin{align}\label{eqn:Cheb1st_temp}
{\partial T_k\circ\psi_{[\tilde{x}_k, \tilde{x}_{k+1}]}(\tilde{x}) \over \partial \tilde{x}}={2k\over d-c}{T_{k+1}\circ\psi_{[\tilde{x}_k, \tilde{x}_{k+1}]}(\tilde{x})-T_{k-1}\circ\psi_{[\tilde{x}_k, \tilde{x}_{k+1}]}(\tilde{x})\over 1-\tilde{x}^2 }.
\end{align}

In a similar fashion, we can obtain $\Gamma_t$ by differentiating $\Delta_t$ with respect to $S$ such that 
\begin{align}\label{eqn:CFS_gamma}
\Gamma_t&={\partial^2 V(x,K,t) \over \partial S^2}={\partial \Delta_t \over \partial S}={\partial \Delta_t \over \partial x}{\partial x \over \partial S}\nonumber\\
&=e^{-r(T-t)-2x}K\left({\partial^2 h(\tilde{x}) \over \partial^2 \tilde{x}}-{\partial h(\tilde{x}) \over \partial \tilde{x}}\right)\nonumber\\
&=e^{-r(T-t)-2x}K\Big(\sum\limits_{k=1}^{M_k} {\partial^2 V_{N_k} \over \partial^2 \tilde{x}} \mathds{1}_{[\tilde{x}_k, \tilde{x}_{k+1}]} -\sum\limits_{k=1}^{M_k}{\partial V_{N_k} \over \partial \tilde{x}} \mathds{1}_{[\tilde{x}_k, \tilde{x}_{k+1}]}\Big).
\end{align}
To find $\partial^2 V_{N_k}/\partial^2 \tilde{x},$
 we may use $${d^2\over d x^2}T_n(x)={n\over 4}{(n+1)T_{n-2}(x)-2nT_n(x)+(n-1)T_{n+2}(x)\over (1-x^2)^2}$$ \citep[cf.][Problem 2.5.17]{Mas_Han:2002}, and thus find
\begin{align}
{\partial^2 V_{N_k} \over \partial^2 \tilde{x}} =\sum\limits_{k=1}^{N_k}\gamma_k{\partial^2 T_k\circ\psi_{[\tilde{x}_k, \tilde{x}_{k+1}]}(\tilde{x}) \over \partial^2 \tilde{x}}
\end{align}
and
\begin{align}
{\partial^2 T_k\circ\psi_{[\tilde{x}_k, \tilde{x}_{k+1}]}(\tilde{x}) \over \partial^2 \tilde{x}}&={k\over (d-c)(1-\tilde{x}^2)^2}\Bigg({(k+1)T_{k-2}\circ\psi_{[\tilde{x}_k, \tilde{x}_{k+1}]}(\tilde{x})}\nonumber\\
&\quad-{2kT_{k}\circ\psi_{[\tilde{x}_k, \tilde{x}_{k+1}]}(\tilde{x})+(k-1)T_{k+2}\circ\psi_{[\tilde{x}_k, \tilde{x}_{k+1}]}(\tilde{x})}\Bigg).
\end{align}

Likewise, we can obtain the formula for Vega, ${\partial V\over \partial \sigma_t},$ where $\sigma_t$ is the initial value of the volatility at time $t.$ For example, for the GBM model with $\sigma_t$ as the initial value of the volatility, we derive Vega as follows:
\begin{align}\label{eqn:vega}
{\partial V(x,K,\sigma_t, t)\over \partial \sigma_t} &=e^{-r(T-t)}K\int_{\min(d,\tilde{x}-c)}^{\max(c,\tilde{x}-d)} f(y){\partial g^R(\tilde{x}-y)\over \partial \sigma_t} \mathrm{d}y,\quad \tilde{x}\in[2c,2d].
\end{align} 
After we differentiate $V$ with respect to $\sigma_t$ to obtain (\ref{eqn:vega}), we can approximate (\ref{eqn:vega}) with the CONLeg method. 

If the closed-formed PDF $g$ of the stochastic process does not exist, as we mentioned before, we express $g^R$ with the CFS expression such that 
\begin{align}
{\partial g^R(\tilde{x}-y)\over \partial \sigma_t}=\mathfrak{Re}\left[2\sum_{k=1}^{N} {\partial b_k \over \partial \sigma_t}e^{-i\frac{2\pi}{d-c}k(\tilde{x}-y)}\right],\hbox{ and } {\partial b_k\over \partial \sigma_t}={\partial \varphi(-\frac{2\pi}{d-c}k, \sigma_t)\over \partial \sigma_t},
\end{align} 
where $\varphi$ contains the parameter $\sigma_t.$ 

\section{Choice of Truncated Intervals}\label{sec:trunc}
In this section, we adopt the ideas of \citet{Fan_Oos:2009a} and \citet{Chan:2018} to choose the interval $[c,d]$. The choice of the interval $[c,d]$ plays the crucial role in the accuracy of the CONLeg method. If the choice of $[c,d]$ is too big, the CONLeg method can perform inefficiently. On the contrast, if $[c,d]$ is too small, the CONLeg method can produce inaccurate option prices/hedging values. Accordingly, a minimum and substantial interval $[c,d]$ can be chosen to capture most of the mass of a PDF such that our algorithm can in turn yield the highest accuracy. In this short section, we show how to construct an interval related to the closed-form formulas of stochastic process cumulants. The idea of using the cumulants is first proposed by \citet{Fan_Oos:2009a} to construct the definite interval $[c,d]$ in (\ref{eqn:charAppx}). Based on their ideas, we have the following expression for $[c,d]$:
\begin{align}\label{eqn:truncInt1}
d&= \left\vert c_1+L_n\sqrt{c_2+\sqrt{c_4}} \right\vert\nonumber\\
c&=-d,
\end{align}
where $c_1,$ $c_2,$ and $c_4$ are the first, second and fourth cumulants, respectively, of the stochastic process and $L_n\in[8,12].$ For simple, less-complicated financial models, we also obtain closed-form formulas for $c_1,$ $c_2,$ and $c_4$, which are shown in Table \ref{table:cumulants}.

\begin{table}[h]
\caption{The first, second, and fourth cumulants of various models.} \label{table:cumulants} 
\centering 
\small\addtolength{\tabcolsep}{-1pt}
\begin{tabular}{|l|l|} 
\hline
\multicolumn{1}{|l|}{L\'evy models}&cumulants\\
\hline 
BS&$c_1=(r-q+\omega)t$\quad$c_2=\sigma^2 t,$\quad $c_4=0,$ $\omega=-0.5\sigma^2$\\
\hline
NIG&$c_1=(r-q+\omega)t+\delta t\beta/\sqrt{\alpha^2-\beta^2}$\\
\,&$c_2=\delta t \alpha^2 (\alpha^2-\beta^2)^{-3/2}$\\
\,&$c_4=\delta t \alpha^2 (\alpha^2+4\beta^2)^{-3/2}(\alpha^2-\beta^2)^{-7/2}$\\
\,&$\omega=-0.5\sigma^2-\delta(\sqrt{\alpha^2-\beta^2}-\sqrt{\alpha^2-(\beta+1)^2}) $\\
\hline 
VG&$c_1=(r-q+\theta+\omega)t$\\
\,&$c_2=(\sigma^2+\upsilon\theta^2)t$\\
\,&$c_4=3(\sigma^4\upsilon+2\theta^4\upsilon^3+4\sigma^2\theta^2\upsilon^2)t$\\
\,&$\omega={1 / \upsilon}\log(1-\theta\upsilon-\sigma^2\upsilon/2)$\\
\hline 
$\rm CGMY$&$c_1=(r-q+\omega)t$\\
\,&$c_2=(C\Gamma(2-Y)(M^{Y-2} + G^{Y-2})t$\\
\,&$c_4=(C\Gamma(4-Y)(M^{Y-4} + G^{Y-4})t$\\
\,&$\omega=\left(C\Gamma(-Y)G^Y\left(\left(1+\frac{1}{G}\right)^Y-1-\frac{Y}{G}\right)+C\Gamma(-Y)M^Y\left(\left(1-\frac{1}{M}\right)^Y-1+\frac{Y}{M}\right)\right)$\\
\hline 
\end{tabular}
\end{table}

In general, the truncated intervals in (\ref{eqn:truncInt1}) work for smooth/non-smooth PDFs with/without singularities. However, as the cumulants in each process in Table \ref{table:cumulants} contain $t$ , we notice that if $t$ is too small, the cumulants shrink and make $[c,d]$ too small for the CONLeg to produce accurate option prices/hedging values. To solve the problem heuristically, as long as $t$ is less than or equal to $0.2,$ we should add extra 0.5 to $d$ in (\ref{eqn:truncInt1}) in order to produce a substantial interval $[c,d]$ for the CONLeg method. 

\section{Numerical Results}\label{sec:results}
The main purpose of this section is to test the accuracy and efficiency of the CONLeg method through various numerical tests. This involves the ability of the method to price any options that are deep in/out of the money and have long/short maturities. Most importantly, we show that the algorithm exhibits good accuracy even when the PDF is smooth/non-smooth. A number of popular numerical methods are implemented to compare the algorithm in terms of the error convergence and computational time. These methods include the COS method (a Fourier COS series method, \citealp{Fan_Oos:2009a}), the filter-COS method (a COS method with an exponential filter to resolve the Gibbs phenomenon; see \citealp{Ruj_Oos:2013}), the CONV method (an FFT method, \citealp{Lor_Fan:2008}), the Lewis-FRFT (a fractional FFT method, \citealp{Lew:2001, Cho:2004}), the QUAD-CONV (a combination of the quadrature and CON methods; see \citealp{Sul:2005, Che_New:2014}), and the SWIFT methods (a wavelet-based method; see \citealp{Gra_Oos:2013, Mar:2015, Gra_Oos:2016, Mar_Gra:2017}). When we implement the CONV and  Lewis-FRFT methods, we use Simpson's rule for the Fourier integrals to achieve fourth-order accuracy. In the filter-COS method, we use an exponential filter and set the accuracy parameter to $10$ as \cite{Ruj_Oos:2013} report that this filter provides better algebraic convergence than the other options. We also set the damping factors of the CONV to 0 for pricing European options. A MacBook Pro with a 2.8 GHz Intel Core i7 CPU and two 8 GB DDR SDRAM (cache memory) is used for all experiments. Finally, the code is written in MATLAB and also the codes of implementing the COS method and the FFT method, such as the CONV method and the like, is retrieved from \citet{BEN:2015}.

Since we use the built-in Chebfun \citep{Tre_Dis:2014} commands, mainly {\tt chebfun, conv, diff, roots and simplify}, to price and hedge the options in this paper, Chebfun (\url{http://www.chebfun.org/download/}) is required for our option pricing algorithm. For a demonstration, we give the MATLAB code for computing European options in Appendix \ref{codes:Ber}. As Chebfun makes use of adaptive procedures that aim to find the right number of points automatically so as to represent each function to roughly machine precision, that is, about 15 digits of relative accuracy, we allow Chebfun to approximate $f$ and $g^R$ automatically. As Chebfun is implemented in MATLAB, we should point out that since Chebfun calls functions, nested functions and sub-functions considerably, the overheads occur and accordingly, this slows the CONLeg method's computation speed down. Hence, as an introductory of the CONLeg method, we weigh more on the convenience of the method over its efficiency at this stage of development. 

In all numerical experiments,  the range of option prices we measure is based on the input range of $S$ or $K.$ To allow our method to have accurate approximation of deep in/out-of-money option prices, the range of $S$ or $K$ lie in the intervals $[K-20, K+20]$ or $[S-20, S+20]$ respectively. Moreover, as we look at the intervals of $[K-20, K+20]$ or $[S-20, S+20],$ we set $[c,d]$ rather than $[2c,2d]$ for approximate option prices and their option Greek values \footnote{To achieve this, we can set a flag--\enquote{same}--in {\tt conv}. See Appendix \ref{codes:Ber} for details.}.  

We use $N$ to define the number of terms/grid points of the methods we compare against the CONLeg method. When measuring approximation errors of the numerical methods, we use absolute errors, the $L_2$ norm errors $R_2$ and the infinity norm errors $R_\infty$ as the measurement units. Moreover, to improve the accuracy of our method in pricing/hedging European type options, we use the well-known call-put parity relationship,
\begin{align}\label{eqn:callputparity}
V_{call}(x,K,t) = V_{put}(x,K,t) + Se^{-q(T-t)}-Ke^{-r(T-t)},
\end{align}
to approximate call prices once we have put prices ready. 

\subsection{European Type Options}\label{sec:results_Euro}
We consider three different test cases based on the following PDFs and other parameters:
\begin{flalign}
\textbf{GBM1}:\,S&= 100,\, K = 80-120,\, \sigma=0.15,\, T = 1.0,\, r= 0.03,\, q= 0.01. \label{BSM1}\\ 
\textbf{GBM2}:\, S&= 100,\, K = 80-120,\, \sigma= 0.25,\, T= 50 \hbox{ or } 100,\, r = 0.1,\, q = 0.\\
\textbf{VG1}:\, S&=100,K = 80-90,\, \sigma=0.12,\, \theta=-0.14,\,\nu=0.2,\,T=0.1,\, r=0.1,\, q=0.
\end{flalign}

In all three numerical tests, the reference values for the GBM process and the VG process are generated via MATLAB Financial Toolbox\texttrademark--{\tt blsprice, blsdelta and blsgamma}--and the Singularity Fourier--Pad\'e (SFP) method \citep[cf.][]{Chan:2018} respectively. We also set $L_n=10$ in (\ref{eqn:truncInt1}) when we create a computational interval $[c,d]$ for pricing/hedging European options under the two processes.

In the first numerical test (\textbf{GBM1})--Table \ref{table:BS_normal_CongLeg}, we first check for convergence behaviour against a range of strikes $K$ from $80$ to $120$ for deep in/out-of-the money and at-the-money vanilla put options. Apart from $q= 0.01,$ the parameters are retrieved from \citet{BEN:2015}. We declare 1000 different option prices within the range of either $K.$ In this test, even without applying the put-call parity (\ref{eqn:callputparity}), the CONLeg can achieve very high accuracy (around $R_\infty=10^{-14}$) when it is applied to approximate option prices and its Delta $\Delta$ and Gamma $\Gamma.$ Moreover, since our method aims to model option price/Greek curves rather than their values, our method consumes only less than 0.1 seconds to formulate the curves in the test. Using around 0.1 seconds to produce option price and Greek curves is a quite reasonable computational cost for a method to meet the financial standards. The second numerical experiment (\textbf{GBM2}) is devoted to the performance of the CONLeg method for long maturity call options, which are often encountered in the insurance and pension industry. The parameters are retrieved from \cite{Gra_Oos:2016} for the test. Table \ref{table:BS_LongMa_SFP} refers to the second test (\textbf{GBM2}) and replicates Table 3 in \cite{Gra_Oos:2016}. In this test, with the help of (\ref{eqn:callputparity}), the CONLeg method impressively provides high accuracy when we declare 1000 different option prices within the range of $K$ from 80 to 120. The last set of parameters (\textbf{VG1}) is chosen in the last numerical test (Table \ref{table:VG_ConLeg_individuals}) because relatively slow convergence was reported for the CONV method for very short maturities in \citet{Lor_Fan:2008}. This is attributed to the PDF of the process being sharp-peaked with a difficult logarithmic singularity $x_{sing}$ (see Fig. \ref{fig:VGsingularity}). Before we approximate the closed-form VG PDF defined in (\ref{eqn:VG_PDF}) with Chebyshev series, we apply the Fourier--Pad\'e method (cf. Appendix \ref{sec:sing}) to locate the singularity in the PDF. We use $x_{sing}$ as a breakpoint and approximate the PDF in two different regions $[c,x_{sing}]$ and $[x_{sing},d]$ to resolve Gibbs phenomena. Since we use the Fourier--Pad\'e method to approximate the PDF, the number of Fourier terms in the Fouier--Pad\'e approximation is set to be 1024. In Table \ref{table:VG_ConLeg_individuals}, again with the aid of (\ref{eqn:callputparity}), within similar CPU time and 30 options measured within the range of 80 and 90, the CONLeg method yields almost same accuracy to the COS and filter-COS methods, but relatively higher accuracy than the Lewis-FRFT, QUAD-CONV and CONV methods.

\begin{remark}\label{remark:VG}
\citet{Fan_Oos:2009a} suggest that the VG process with \textbf{VG1} gives rise to a probability density function that is not in $C^{\infty}(\mathds{R})$, and thus, option pricing under VG with these parameter sets exhibits only an algebraic convergence. Nevertheless, \citet{Chan:2018} has recently proposed the SFP method to circumvent the problem and to approximate the VG PDF with the input parameters of \textbf{VG1}. Through this method, we can regain global spectral convergence away from singularities. For more details, we refer the readers to \citet{Chan:2018}. 
\end{remark}
\begin{table}
\center
\caption{Measuring the CONLeg method in error convergence and CPU time for pricing European at/around-the-money put prices, Delta $\Delta_t$ and Gamma $\Gamma_t$ under the BSM model with parameters taken from \textbf{GBM1}. 1000 put prices are computed in a range of $K$ from 80 to 120.}
\label{table:BS_normal_CongLeg}
\small\addtolength{\tabcolsep}{-1.5pt}
\begin{tabular}{|cc|cc|cc|c|} 
\hline
\multicolumn{2}{|c}{Price}& \multicolumn{2}{|c}{Delta\,$\Delta_t$}& \multicolumn{2}{|c|}{Gamma\,$\Gamma_t$}&\\
$R_\infty$&$R_2$&$R_\infty$&$R_2$&$R_\infty$&$R_2$&Time\\
\hline
5.329e-14 &   3.921e-13 &   5.645e-14  &  1.846e-13  & 8.538e-13  &  1.396e-12&9.11e-02\\
\hline
\end{tabular}
\end{table}

\begin{table} 
\caption{Measuring the CONLeg method in error convergence and CPU time for pricing European at/around-the-money call prices, Delta $\Delta_t$ and Gamma $\Gamma_t$ under the BSM model with parameters taken from \textbf{GBM2}. 1000 call prices are computed in a range of $K$ from 80 to 120.} 
\label{table:BS_LongMa_SFP}
\centering 
\small\addtolength{\tabcolsep}{-1.5pt}
\begin{tabular}{|c|cc|cc|cc|c|} 
\hline
&\multicolumn{2}{|c}{Price}& \multicolumn{2}{|c}{Delta\,$\Delta_t$}& \multicolumn{2}{|c|}{Gamma\,$\Gamma_t$}&\\
&$R_\infty$&$R_2$&$R_\infty$&$R_2$&$R_\infty$&$R_2$&Time\\
\hline
$T=50$&2.842e-14&    2.821e-13&    2.220e-16 &   1.005e-15 &   7.170e-17&    5.007e-16&8.34e-02\\
\hline
\multicolumn{8}{c}{}\\
\hline
&$R_\infty$&$R_2$&$R_\infty$&$R_2$&$R_\infty$&$R_2$&Time\\
\hline
$T=100$&2.842e-14&    2.874e-13&  3.841e-16&    2.123e-16&  6.126e-19&    3.776e-18&9.12e-02\\
\hline
\end{tabular}
\end{table}


\begin{table}
\caption{Comparison of the Lewis-FRFT, CONV, QUAD--CONV, COS, filter-COS and CONLeg methods for pricing vanilla call options under the VG model with parameters taken from \textbf{VG1}. 30 call prices are computed in a range of $K$ from 80 to 90.} \label{table:VG_ConLeg_individuals}
\centering 
\small\addtolength{\tabcolsep}{-1.5pt}
\begin{tabular}{|c|ccc|ccc|ccc|} 
\hline
&\multicolumn{3}{c}{\textbf{Lewis-FRFT}}&\multicolumn{3}{|c}{\textbf{CONV}}&\multicolumn{3}{|c|}{\textbf{QUAD--CONV}}\\
\hline
$N$&$R_\infty$&$R_2$&Time&$R_\infty$&$R_2$&Time&$R_\infty$&$R_2$&Time\\
1024&9.921e-03 &  1.121e-02 & 0.092 &  1.217e-04 &   8.817e-03  & 0.110& 2.411e-04 &   7.827e-03& 0.121\\
\hline
\multicolumn{10}{c}{}\\
\hline
&\multicolumn{3}{c}{\textbf{COS}}&\multicolumn{3}{|c|}{\textbf{filter-COS}}&\multicolumn{3}{c|}{\textbf{CONLeg}}\\
\hline
$N$&$R_\infty$&$R_2$&Time&$R_\infty$&$R_2$&Time&$R_\infty$&$R_2$&Time\\
1024&2.271e-06& 7.579e-05& 0.089&  6.596e-07&  3.596e-06  & 0.0912 & 5.596e-07&  4.511e-06  & 0.102\\
\hline
\end{tabular}
\end{table}
\subsection{American, Barrier and Bermuda Options}\label{sec:results_Exotic}
In this section we again focus on three test cases based on the following PDFs and other parameters:
\begin{align}
\textbf{NIG1}:\, S&=100,\, K=80-120,\, \alpha=15,\, \beta=-5,\, \delta=0.5,\, T=1,\, r=0.05,\, q=0.02.\\
\textbf{CGMY1}:\, S&=0.5-1.5,\, K=1,\, C=1,\, G=5,\, M=5,\, Y=0.5,\, T=1,\, r=0.1,\, q=0.0.\\
\textbf{CGMY2}:\, S&=90-100,\, K=100,\, C=4,\, G=50,\, M=60,\, Y=0.7,\, T=1,\, r=0.05,\, q=0.02.
\end{align}
In this section, as suggested by \citet{Fan_Oos:2009b}, we first set $L_n=8$ in (\ref{eqn:truncInt1}) to create a computational interval $[c,d]$ for approximating the PDFs of the NIG and CGMY processes. In our first test we price Bermudan put options with 10 exercise dates with parameters of \textbf{NIG1}. In this test (Table \ref{table:NIG_ConLeg_170}), a total of 170 option prices are generated in the COS and CONV methods from a range of $K$ between 80 to 120. The CPU times are reported in seconds, and all reference values are obtained by the CONV method with $N = 2^{20}.$ \citet{Fan_Oos:2009b} report that the NIG PDF (\ref{eqn:NIG_PDF}) is more peaked at the mean with the parameters of \textbf{NIG1}. We then use the mean as a breakpoint to approximate the PDF in two regions $[c,x_{sing}]$ and $[x_{sing},d].$  From the result of Table \ref{table:NIG_ConLeg_170}, the accuracy of the COS method stays the same when $N$ increases from 256 to 512. Within the similar CPU time, we can see that the CONLeg has better accuracy than the COS method. 

In the next two tests, the CGMY processes with parameters of \textbf{CGMY1} and \textbf{CGMY2} do not have a closed-form PDF and have singularities. Accordingly, again using the techniques we mentioned above, we use the Fourier--Pad\'e method to approximate the CGMY PDFs and locate their singularities. The number of Fourier terms in the Fouier--Pad\'e approximation is set to be 1024 in the tests.

The prices of American options can be obtained using (\ref{eqn:amer_extrapolation}), a 4-point Richardson extrapolation method on the prices of a few Bermudan options with small $L.$ We compare the CONV, COS and CONLeg methods for pricing American option price in Table \ref{table:CGMY_ConLeg_Amer}. All reference values are obtained by the Fourier time stepping method \citep{Jac_Jai_Sur:2008} with the number of the Fourier terms equal to $2^{16}.$ Within the similar CPU time, the CONLeg method achieves the same accuracy with the COS method and is marginally better than the CONV method.

For the last two tables, all reference values are obtained by the CONV method with $N = 2^{20}.$ In Table \ref{table:CGMYNIG_ConLeg_Barrier}, we consider monthly monitored $(L= 12)$ up-and-out call and put options, (UO Call) and (UO Put) under the CGMY process with the parameters of \textbf{CGMY2}. The barrier level $H$ is set to be 120 for the up-and-out options. As we can see from the table, the CONLeg method is very comparable to the SWIFT method in terms of accuracy. Finally, in Table \ref{table:NIG252_ConLeg_Barrier}, we focus on the NIG process with parameters of \textbf{NIG1}, except S=90--110 and K=100. Considering monthly monitored $(L= 252)$ down-and-out call (DO Call) and put (DO Put) options, we set $H$ equal to 80 for the options. Since $L=252$ is very large, it causes the time interval difference between $t_{l+1}$ and $t_{l}$ to be very small. Accordingly, the NIG PDF is very peaked and the COS method requires a larger number of $N$ equal to 8192 to compensate a better convergence in Table \ref{table:NIG252_ConLeg_Barrier}. Comparing the CONLeg and COS methods in the similar CPU time, both methods can achieve similar accuracy.
\begin{table}
\caption{Comparison of the COS and CONLeg methods for pricing a Bermuda ($L = 10$) put option under the NIG model with parameters taken from \textbf{NIG1}. 170 option prices are generated for both methods in a range of $K$ from 80 to 120 and $S$ equal to 100.} \label{table:NIG_ConLeg_170}
\centering 
\begin{tabular}{|c|ccc|ccc|} 
\hline
&\multicolumn{3}{c}{\textbf{COS}}&\multicolumn{3}{|c|}{\textbf{CONLeg}}\\
$N$&$R_\infty$&$R_2$&Time&$R_\infty$&$R_2$&Time\\
\hline
256& 2.411e-07&5.411e-06&2.89&1.812e-13&8.576e-12&2.93\\
\hline
\multicolumn{7}{c}{}\\
\hline
$N$&$R_\infty$&$R_2$&Time&$R_\infty$&$R_2$&Time\\
\hline
512& 2.411e-07&5.411e-06&3.323&1.812e-13&8.576e-12&2.93\\
\hline
\end{tabular}
\end{table}


\begin{table}
\caption{Comparison of the CONV, COS and CONLeg methods for pricing an American put option under the CGMY model with parameters taken from \textbf{CGMY1}. 58 and 88 option prices are computed for the CONV method and the COS method respectively in a range of $S$ from 0.5 to 1.5 and $K$ equal to 1. } \label{table:CGMY_ConLeg_Amer}
\centering 
\begin{tabular}{|c|c|ccc|ccc|} 
\hline
\multirow{2}{*}{$L$ in Eq. (\ref{eqn:amer_extrapolation})}&\multirow{2}{*}{$$}&\multicolumn{3}{c|}{\textbf{CONV}}&\multicolumn{3}{c|}{\textbf{CONLeg}}\\
&$N$&$R_\infty$&$R_2$&Time (sec.)&$R_\infty$&$R_2$&Time (sec.)\\
\cline{2-8}
2&32768& 1.052e-04&8.051e-03&19.291&7.123e-05&6.113e-04&19.311\\
\hline
\multicolumn{8}{c}{}\\
\hline
\multirow{2}{*}{$L$ in Eq. (\ref{eqn:amer_extrapolation})}&\multirow{2}{*}{$$}&\multicolumn{3}{c|}{\textbf{COS}}&\multicolumn{3}{c|}{\textbf{CONLeg}}\\
&$N$&$R_\infty$&$R_2$&Time (sec.)&$R_\infty$&$R_2$&Time (sec.)\\
\cline{2-8}
2&32768& 9.234e-05&4.012e-04&19.298&7.123e-05&6.113e-04&19.311\\
\hline
\end{tabular}
\end{table}
\begin{table}
\caption{Comparison of the SWIFT and CONLeg methods for pricing daily-monitored ($L = 12$) UO call and UO put under the CGMY model with parameters taken from \textbf{CGMY2}. 725 option prices are computed in a range of $S$ from 90 to 110 and $K$ equal to 100. The barrier level $H$ is equal to 120.} \label{table:CGMYNIG_ConLeg_Barrier}
\centering 
\begin{tabular}{|c|cccc|ccc|} 
\hline
&\multicolumn{4}{c|}{\textbf{SWIFT}}&\multicolumn{3}{c|}{\textbf{CONLeg}}\\
&$scale$&$R_\infty$&$R_2$&Time (sec.)&$R_\infty$&$R_2$&Time (sec.)\\
\hline
UO Call&6&3.181e-10&7.641e-09&7.100&5.186e-09&6.886e-08&7.201\\
\hline
UO Put&6&1.901e-11&8.621e-10&6.901&4.330e-10&6.131e-09&6.891\\
\hline
\end{tabular}
\end{table}

\begin{table}
\caption{Comparison of the COS and CONLeg methods for pricing daily-monitored ($L = 252$) DO Call and DO Put under the NIG model with parameters taken from \textbf{NIG1}. 38 option prices are computed in the range of $S$ from 90 to 110 and $K$ equal to 100. The barrier level $H$ is equal to 80.} \label{table:NIG252_ConLeg_Barrier}
\centering 
\begin{tabular}{|c|cccc|ccc|} 
\hline
&\multicolumn{4}{c|}{\textbf{COS}}&\multicolumn{3}{c|}{\textbf{CONLeg}}\\
&$N$&$R_\infty$&$R_2$&Time (sec.)&$R_\infty$&$R_2$&Time (sec.) \\
\hline
DO Call&8192&6.701e-08&4.641e-07&94.231&4.167e-09&8.651e-08&93.812\\
\hline
DO Put&8192&7.232e-09&3.231e-08&92.532&1.619e-09&6.886e-08&92.634\\
\hline
\end{tabular}
\end{table}

\section{Conclusions}\label{sec:conclusion}
In this paper, we have proposed an algorithm for hedging and pricing various options based on approximating probability density functions by polynomials (in particular, Legendre series) and computing the prices by performing suitable convolutions with the given payoff function. We call this method CONLeg. The main advantages of the CONLeg method are its ability to return the price and Greeks as a function defined on a prescribed interval rather than just point values, its ability to approximate different types options under a process with/without a closed-from PDF, and the simplicity and convenience of its implementation using Chebfun. We presented a proof of concept implementation written in Chebfun, which demonstrates positive results in the accuracy of the method. Since Chebfun uses adaptive approximations, it is difficult at this time to discuss rates of convergence of the CONLeg method or to compare it's efficiency with respect to other methods in the literature (which work with a fixed discretisation size), but initial results are promising. For a fair comparison, one would need to implement a non-adaptive version of CONLeg, which would be time-consuming, and not typically desirable in a practical setting.

Our ultimate goal is to extend the method to price options with path-dependant features under the (time-changed) L\'evy process or stochastic volatility with and without singularities. Research in this direction is already underway and will be presented in a forthcoming manuscript.

\begin{APPENDICES}
\section{A Closed-form Transformation of Complex Fourier Series into Chebyshev Series}\label{sec:CFS_Cheb}
To transform a complex Fourier series (CFS) into a Chebyshev series, we apply the result of \citet[Lemma A.3]{Tow:2014}, i.e.
\begin{align}\label{eqn:exp_J}
\int_{-1}^1{\exp(ixy\pi)T_q(x)\over\sqrt{1-x^2}}\mathrm{d}x=\pi i^qJ_q(y\pi).
\end{align}
Here, $J_q$ is the Bessel functions of the first kind with parameter $q$ and $T_q(x)$ is the Chebyshev polynomial of degree $q.$ We first note that since $g,$ a PDF, is a real function, the CFS representation of $g$ with a form of 
\begin{align}
g(x)&\approx g_N(x)=\mathfrak{Re}\left[2\sum_{k=1}^{N}\varphi\left(-\frac{2\pi}{d-c}k\right) e^{i\frac{2\pi}{d-c}kx}+\varphi(0)\right]. 
\end{align}
can be interchangable into 
\begin{align}\label{eqn:CFS_PDFN}
\mathfrak{Re}\left[ \sum_{k=-N}^N \varphi\left(-\frac{2\pi}{d-c}k\right) e^{i{2\pi\over d-c}kx}\right].
\end{align}
Using (\ref{eqn:exp_J}) and (\ref{eqn:CFS_PDFN}), we may define a Chebyshev series on the same interval $[c,d]$ such that 
$$\mathfrak{Re}\left[\sum_{k=-N}^{N} \varphi\left(-\frac{2\pi}{d-c}k\right) e^{i\frac{2\pi}{d-c}kx}\right]=\sum_{n=0}^N \alpha_n^{cheb} T_n\circ\psi_{[c,d]}(x),$$
where,
\begin{eqnarray}
\alpha^{cheb}_0&=&\mathfrak{Re}\left[\sum_{\substack{k=-N\\k\not=0}}^{N} {\Gamma\left({1\over 2}\right)\over \sqrt{\pi}} \varphi\left(-\frac{2\pi}{d-c}k\right) e^{i\frac{(d+c)}{d-c} k} J_0\left(- k\pi\right)+\varphi(0)\right],\, n=0,\nonumber\\
\alpha^{cheb}_n&=&\mathfrak{Re}\left[2\sum_{\substack{k=-N\\k\not=0}}^{N} \varphi\left(-\frac{2\pi}{d-c}k\right) e^{i\frac{(d+c)}{d-c} k} i^nJ_n\left(- k\pi\right)\right],\, n> 0.\nonumber\\
\end{eqnarray}
\section{Locating Singularities in Probability Density Functions}\label{sec:sing}
Many PDFs of interest are not smooth but piecewise smooth. For example, see Figure \ref{fig:VGsingularity}. If the locations of all singularities are not known in advance, we can use Fourier--Pad\'e ideas \citep[cf.][]{Dris_Ben:2011,Chan:2018} to estimate the locations of singularities well enough to allow good reconstruction nearly everywhere in the interval $[c,d]$. 

The Fourier--Pad\'e algorithm proposed in this paper is very simple to implement. If we consider a function $g$ with a power series representation such that 
$$g(x)=\sum_{k=0}^\infty b_k x^k,$$ 
and a rational function defined by $R_{N,M}=P_N/Q_M,$ where $P_N$ and $Q_M$ are the polynomials of 
\begin{align}\label{eqn:FP_1}
P_N(x)=\sum_{n=0}^N p_{n} x^{n}\hbox{ and } Q_M(x)=\sum_{m=0}^M q_{m} x^{m},
\end{align}
respectively, then we say that $R_{N,M}=P_N/Q_M$ is the \textsl{(linear) Pad\'e approximant} of order (N, M) of the formal series satisfying the condition 
\begin{align}\label{eqn:FP_2}
\left(\sum_{n=0}^N p_{n} x^{n}\right)-\left(\sum_{m=0}^M q_{m} x^{m}\right)\left( \sum^{M+N}_{k=0} b_k x^k\right)&=\mathcal{O}(x^{N+M+1}).
\end{align}
Here, $g$ is approximated by $\sum^{M+N}_{k=0} b_k x^k,$ To obtain the approximant $R(N, M),$ we simply calculate the coefficients of polynomials $P_N$ and $Q_M$ by solving a system of linear equations. To obtain $\{q_m\}_{m=0}^M,$ we first normalise $q_0 = 1$ to ensure that the system is well determined and has a unique solution in (\ref{eqn:FP_2}). Then, we consider the coefficients for $x^{N+1},\ldots , x^{M+N},$ and we can yield a Toeplitz*\footnote{A Toeplitz matrix or diagonal-constant matrix is an invertible matrix in which each descending diagonal from left to right is constant.} linear system:
\begin{align}
\begin{bmatrix} 
b_{N+1} & b_{N} & b_{N-1} & \cdots & b_{N+1-M}\\ 
b_{N+2} & b_{N+1} & b_{N} & \ddots & b_{N+2-M}\\ 
\vdots&\ddots &\ddots &\ddots &\vdots\\
b_{N+M} & \cdots &b_{N+2} & b_{N+1} & b_{N}
\end{bmatrix}
\begin{bmatrix}
q_0\\q_1\\ \vdots\\ q_M
\end{bmatrix}
=0.
\end{align}

Once $\{q_m\}_{m=0}^M$ is known, $\{p_n\}_{n=0}^N$ is found through the terms of
order N and less in (\ref{eqn:FP_2}). This yields $\underline{p} = B\underline{q}$, where $b_{ij} = b_{i-j}$. For example, if
$N=M,$ one obtains
\begin{align}
\begin{bmatrix}
p_0\\p_1\\ \vdots\\ p_N
\end{bmatrix}=\begin{bmatrix} 
b_{0} & & & \\ 
b_{1} & b_{0} & & \\ 
\vdots&\ddots &\ddots &\\
b_{N} & \cdots & b_{1} & b_{0}
\end{bmatrix}
\begin{bmatrix}
q_0\\q_1\\ \vdots\\ q_M
\end{bmatrix}.
\end{align}

Now, assuming $g$ is a PDF, to find singularities in $g$ and to express $g$ in a Fourier--Pad\'e series, we first express $g$ with the CFS representation:

\begin{align}\label{eqn:CFS_PDF_temp2}
\mathfrak{Re}\left[2\sum_{k=1}^{\infty} \varphi\left(-\frac{2\pi}{d-c}k\right) e^{i\frac{2\pi}{d-c}kx}+\varphi\left(0\right)\right].
\end{align}
Then, we can differentiate (\ref{eqn:CFS_PDF_temp2}) with respect to $x$ to obtain 

\begin{align}
\mathfrak{Re}\left[2\sum_{k=1}^{\infty} \left(i\frac{2\pi}{d-c}k\right) \varphi\left(-\frac{2\pi}{d-c}k\right) e^{i\frac{2\pi}{d-c}kx}\right]. 
\end{align}
Finally, we let $z=\exp\left({i\frac{2\pi}{d-c}x}\right)$ in the two equations above, and they are ready for the Fourier--Pad\'e approximation. In general, when the PDF has a singularity, the sharp-peaked singularity point will have an enormously large value after differentiation. In other words, Figure \ref{fig:VGsingularity} is a graphical illustration of the outlooks of the PDF (left) and the first derivative (right) of the VG model after the Fourier--Pad\'e approximation. In the figure, we can see that the non-smooth PDF with a jump can produces a value of $10\times 10^{11}$ at the singularity point after the first derivative. See {\tt padeapprox} in Chebfun.
\begin{figure}
\center
\includegraphics[height=5cm,width=10cm]{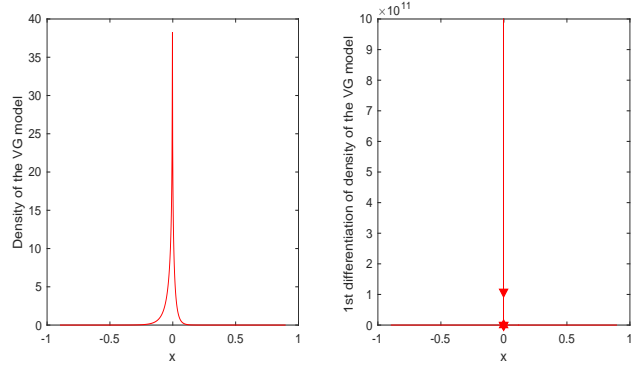}
\setlength{\abovecaptionskip}{1pt}
\caption{Density functions (left) and its first derivative (right) of the VG model with parameters are taken from \textbf{VG1}.}
\label{fig:VGsingularity}
\end{figure}

\section{MATLAB Code}\label{codes:Ber}
\begin{lstlisting}
% COMPUTE GBM EUROPEAN OPTION PRICE AND GREEKS USING CONLEG METHOD.
% Tat Lung (Ron) Chan & Nick Hale, Janauary 2019

K = 100;                                      % Strike price
window = [80, 120];                           % Current stock price window
x = log(linspace(window(1), window(2), 5)/K); % Example stock prices
r = 0.035; t = 0.5; q = 0; sigma = 0.2;         % Input parameters
call = true;                                  % (call = false ==> put)
mu = (r-q-0.5*sigma^2)*t;             
L = 10;
c1 = mu; c2 = sigma^2*t; c4 = 0;      % Cumulants (see section 5)
d = abs(c1+L*sqrt(c2 + sqrt(c4)));    % x-variable domain (see section 5)
c = -d; dom = [c, 0, d];

% Payoff function:
if ( call ) 
    payoff = @(x) K*max(exp(x)-1, 0); % Call payoff
else
    payoff = @(x) K*max(1-exp(x), 0); % Put payoff
end
payoff = chebfun(payoff, dom);        % Chebfun representation:

% Guassian Density Function:
mypdf = @(u) 1./sqrt(2*pi*sigma^2*t)*exp(-0.5*(-u-mu)^2./(sigma^2*t));
mypdf = chebfun(mypdf, dom);          % Chebfun representation:

% Continuous stock price:
stock = chebfun(@(x) K*exp(x), dom);

% Compute European option price via convolution:
price = exp(-r*t)*conv(payoff, mypdf, 'same');
price = simplify(price);

% Compute Greeks:
dP = diff(price);      % 1st derivative
delta = dP./stock;
dP2T = diff(price, 2); % 2nd derivative
gamma = (dP2T-dP)./(stock.^2);

%Display figures:
subplot(1,3,1), plot(stock, payoff, stock, price, 'LineWidth', 2); title('Price')
legend('Payoff', 'Price', 'Location', 'NW'), xlabel('S'), 
axis([window, 0, price(x(end))])
subplot(1,3,2), plot(stock, delta, 'LineWidth', 2); title('Delta');
xlabel('S'), xlim(window)
subplot(1,3,3), plot(stock, gamma, 'LineWidth', 2); title('Gamma')
xlabel('S'), xlim(window), shg

% Display table:
Results = [K+0*x ;  stock(x) ; price(x) ; delta(x) ; gamma(x)];
disp(table(Results, 'RowNames', {'strike', 'stock', 'price', 'delta', 'gamma'}))
\end{lstlisting}

\subsection{Output}
\begin{figure}[b]
\center
\includegraphics[height=5cm,width=13cm]{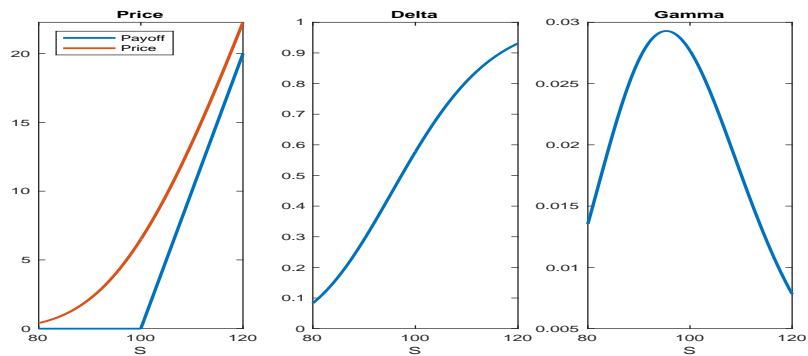}
\caption{Figure produce by code above.}
\end{figure}
\begin{lstlisting}
                                        Results                           
              ____________________________________________________________
    strike         100          100          100          100          100
    stock           80           90          100          110          120
    price       0.4069       2.1637       6.4983      13.5068      22.2810
    delta       0.0833       0.2910       0.5771       0.8074       0.9311
    gamma       0.0135       0.0269       0.0277       0.0176       0.0078

\end{lstlisting}
\end{APPENDICES}



\bibliographystyle{informs2014} 
\bibliography{Q4MMZ42B_AlgoLeg_Nov15} 

\end{document}